\documentclass[12pt,letterpaper]{article} 

\usepackage[utf8]{inputenc} 
\usepackage{float}
\usepackage{subfig}
\usepackage{diagbox}
\usepackage{setspace}
\usepackage{amsmath}        
\usepackage[table]{xcolor}
\usepackage{amssymb}        
\usepackage{graphicx}       
\usepackage{xurl}            
\usepackage{natbib}         
\usepackage{hyperref}       
\usepackage[affil-it]{authblk} 
\usepackage{colortbl}       
\usepackage{booktabs}       

\hypersetup{
    colorlinks=true,
    linkcolor=black,
    filecolor=magenta,
    urlcolor=black,
    citecolor=black,
}
\usepackage[margin=1in]{geometry} 

\usepackage{changepage} 

\pdfoutput=1

\title{\textbf{\Large{Techno-economic analysis of decarbonized backup power systems using scenario-based stochastic optimization}}}

\author[1,2]{Jonas Schweiger} 
\author[1]{Ruaridh Macdonald}

\affil[1]{MIT Energy Initiative, Massachusetts Institute of Technology, Cambridge, MA, United States}
\affil[2]{College of Management of Technology, École Polytechnique Fédérale de Lausanne, Lausanne, Switzerland}

\date{} 

\begin{document}

\maketitle

\subsection*{Abstract}
In the context of growing concerns about power disruptions, grid reliability and the need for decarbonization, this study evaluates a broad range of clean backup power systems (BPSs) to replace traditional emergency diesel generators. A scenario-based stochastic optimization framework using actual load profiles and outage probabilities is proposed to assess the most promising options from a pool of 27 technologies. This framework allows a comparison of cost-effectiveness and environmental impact of individual technologies and hybrid BPSs across various scenarios. The results highlight the trade-off between total annual system cost and emissions. Significant emission reductions can be achieved at moderate cost increases but deep decarbonization levels incur higher costs. Primary and secondary batteries are included in optimal clean fuel-based systems across all decarbonization levels, combining cost-effective power delivery and long-term storage benefits. The findings highlight the often-overlooked importance of fuel replacement on both emissions and costs. Among the assessed technologies, ammonia generators and hydrogen fuel cells combined with secondary iron-air batteries emerge as cost-effective solutions for achieving decarbonization goals. To ensure a broad range of applicability, the study outlines the impact of emergency fuel purchases, varying demand patterns and demand response options on the optimal BPS. The research findings are valuable for optimizing the design of clean BPSs to economically meet the needs of many facility types and decarbonization targets.

\subsection*{Keywords}
Backup power systems; emergency diesel generators; energy system modeling; optimization; decarbonization
\newpage
\section{Introduction}
The use of uninterrupted power supplies during periods of grid instability or complete power outages has increased significantly in recent years. Critical services like healthcare, communication and data centers rely heavily on standalone backup power systems (BPSs). Increasing electricity demand, aging grid infrastructure and more frequent extreme weather events have led to a surge in power outages worldwide and they are expected to continue to become more frequent. \cite{do2023spatiotemporal} 

Today, the majority of backup power demand is met by fossil-fuel powered generators, particularly emergency diesel generators (EDGs). They are a convenient standby power option due to their low upfront cost, quick start up times and safety during operation. The high volumetric energy density of diesel fuel and low capital cost of double wall aboveground fuel tanks \cite{diesel_storage_tanks} make it feasible for EDGs to operate for a week, providing a high degree of resilience for high value and emergency facilities. 

While diesel generators offer numerous benefits, they also have some undeniable downsides. The main concern is their environmental impact. Diesel generators release a significant amount of CO\textsubscript{2} and other pollutants into the atmosphere. Their emission intensity is 1244 kilogram of CO\textsubscript{2} per electrical MWh delivered, accounting for the emissions related to the extraction and processing of diesel fuel. \cite{diesel_emissions_true}  Taking into account other pollutants such as nitrogen oxides (NOx), volatile organic compounds (VOC) and particulate matter (PM), these emissions contribute not only to an increased level of greenhouse gas emissions (GHG) but also to a degradation in air quality.

EDGs also require regular maintenance and fuel replacement which can amount to significant annual expenditures. EDGs have numerous moving parts and complicated fuel injection systems, requiring inspection and testing every one to six months. While EDGs are generally assumed to be reliable, recent studies show that even well-maintained generators have lower reliability than previously assumed. \cite{marqusee2021impact} Contaminated fuel and issues with starting batteries and cooling systems are the most common problems which lead to EDGs not starting when required. \cite{eoff2007diesel} 

Given all these limitations, replacing EDGs is not just an opportunity to decarbonize BPSs but potentially to improve reliability and reduce total maintenance costs. This presents an important opportunity for research to guide the adoption of alternative technologies. 

Most studies of alternative backup power focus on the comparison of individual technologies to EDGs. Biodiesel generators, hydrogen fuel cells and lithium-ion batteries are the most commonly assessed technologies. \cite{yang2024economic} \cite{dimitriou2019fully} \cite{li2018research} However, some research has been conducted on hybrid technology solutions which combine several BPSs technologies. Dimitriou et al. \cite{dimitriou2019fully} proposed a hydrogen-biodiesel-fueled compression ignition internal combustion engine and showed that the carbon dioxide emissions can be significantly reduced but alternative combustion strategies are required at low load conditions. They focused on the technical aspects of BPSs and did not investigate the economic viability of their hybrid system. Sanni et al. \cite{sanni2021analysis} conducted a study on a hybrid system combining solar photovoltaic (PV) generation with diesel and biogas backup which is designed to complement existing EDGs. They included a case-study of the system operating in south-west of Nigeria. Their results show a 61\% decrease in emissions, with a tradeoff between economic and environmental benefits. Faraji et al. \cite{faraji2019comparative} conducted a techno-economic analysis of a PV-plus-battery microgrid for a local clinic. They found that their proposed system led to a significant reduction in the total annual emissions but again required greater upfront investment costs. In their study, Faraji et al. assumed that only critical loads have to be supplied during power outages and optimized their system assuming one 48-hour blackout would occur at a random time each year. Elkholy et al. \cite{elkholy2024techno} discuss a combination of battery energy storage, hydrogen fuel cells and vehicle-to-grid technology as backup power sources. Their study finds that the combination of a variety of backup sources leads to the best cost-effectiveness and reliability. Dong et al. \cite{dong2018battery} investigated the optimal size of a battery and backup generator, highlighting the investment cost trade-off between the two technologies. The authors show that the stochastic outage duration increases total cost of the system and discuss the impact of the average critical customer interruption time.

Researchers have also investigated islanded renewable power systems, which share key characteristics with emergency standby power systems. Gatta et al. \cite{gatta2018replacing} investigated the possibility of a hybrid power plant on Giglio island which could replace diesel generators. They showed that the hybrid power system has both environmental and economic benefits compared to the traditional generator. Pathak et al. \cite{pathak2024utility} proposed a renewable wind, solar PV and battery system capable of providing backup power to local consumers during power outages and demonstrate its technical feasibility. Li et al. \cite{li2020dynamic} investigated the potential of adiabatic compressed air energy storage as backup system for an islanded microgrid operation. Their results suggest that a standalone operation of the system is not sufficient for an uninterrupted power supply and that a combination with an electrical storage device can be a feasible option. Suman et al. \cite{suman2021optimisation} proposed a sustainable microgrid comprised of variable renewable energy (VRE) resources, two generators and a battery located in the Indian state of Bihar. They solved this optimal planning problem using a particle swarm optimization approach and found that the analyzed system can supply 93\% of demand at a high renewable factor. \cite{suman2021optimisation} The inherent uncertainties of power outages necessitates a framework that accounts for uncertainties in the objective function. Stochastic mixed integer linear programming (MILP) models and multi-stage stochastic programming approaches are described in the literature alongside robust optimization techniques. \cite{shen2021optimal} \cite{franco2015robust} \cite{jawad2018robust}


Our study advances the field beyond the literature by providing a comparative analysis of diverse individual and hybrid BPS configurations. We use a comprehensive approach that quantifies the benefits and synergies of various hybrid setups and emerging technologies. To provide a holistic understanding of the trade-off between economic and environmental aspects we introduce a techno-economic analysis to simultaneously evaluate both dimensions. We base our study on real-world data and vary key parameters to ensure our results are applicable to a broader range of applications and scenarios. This approach is formulated using a scenario-based stochastic optimization framework, specifically adapted to account for the probabilistic nature of power outages.

BPSs are used across a wide range of sizes and purposes. This study focuses on systems with a capacity between 500 kW and 50 MW. This range typically serves two key market segments. The first is critical civic infrastructure and essential facilities, such as hospitals, government centers, police stations, and fire departments. In this use case, uninterrupted power supply is crucial for public safety and the maintenance of essential services. The second target application is high-value industry; including pharmaceutical laboratories, data centers, and airports; where power disruptions can lead to significant financial losses and operational disruptions. While combined backup power solutions incorporating heating or cooling exist, the majority of systems are designed primarily for electricity generation. Thus, this study limits its scope to the provision of electricity, allowing it to focus on the functionality and replacement of conventional backup power generators, such as EDGs. This choice is further justified by the anticipated electrification of heat generation in the future.

The principal objective of this study is to assess cost-effective and decarbonized emergency backup power systems for a variety of applications. The study aims to determine the most suitable technologies using a scenario-based stochastic optimization approach and a capacity expansion model. The novelties of the work include a techno-economic analysis of emerging technologies, a comprehensive cost and emission accounting framework, as well as optimized capacity sizing based on historic load profiles and outage probabilities.

\section{Materials and Methods}

The study uses a three-step approach to analyze the feasibility and competitiveness of alternative BPSs. The first step involves the qualitative assessment of a variety of alternative BPS technologies and the characterization of the most promising substitute technologies. In a second step, the characteristics and patterns of power outages are analyzed. This supplies the outage scenarios and probabilities for the subsequent scenario-based stochastic optimization of the BPSs design and operation. This third step is conducted using the GenX capacity expansion model. This scenario-based approach is used to accurately depict the random nature of power outages and enables the evaluation of different technologies under a variety of conditions, such as variations in demand patterns, demand response options and different decarbonization pathways.

\subsection{Qualitative assessment of alternative technologies}

The qualitative assessment of alternative technologies consists of a comparison of 27 technologies, assessing the strengths, limitations and feasibility of each option. The categorization of these technologies is done along the following dimensions: conventional fuels, clean fuels, rechargeable batteries, non-rechargeable batteries, onsite VRE and thermal systems. Rechargeable batteries and non-rechargeable batteries will hereafter be referred to as secondary and primary batteries. The goal is to conduct a comprehensive comparison that reduces the number of technologies to be assessed more thoroughly in the subsequent sections. Table \ref{tab:dimensions_qualitative_assessment} shows the set of criteria that is used to facilitate this assessment.\\

\begin{table}[h!]
\centering
\setlength{\arrayrulewidth}{0.5mm}
\setlength{\tabcolsep}{10pt}
\renewcommand{\arraystretch}{1.5}
\begin{tabular}{p{3cm} p{12cm}}
\hline
\rowcolor[HTML]{FFFFFF} 
\textbf{Dimension} & \textbf{Description} \\ \specialrule{0.2pt}{0pt}{0pt}
\rowcolor[HTML]{FFFFFF} 
Technical & Factors in efficiency, power output, operational lifespan and provides insights into the technology's performance. \\ 
\rowcolor[HTML]{FFFFFF} 
Physical & Inncludes gravimetric and volumetric energy density, storage requirements, and practicality. \\ 
\rowcolor[HTML]{FFFFFF} 
Maintenance & Evaluates ease of maintenance, servicing frequency, and long-term operational costs. \\ 
\rowcolor[HTML]{FFFFFF} 
Safety & Crucial element for on-site generation that involves an assessment of potential hazards, flammability, toxicity and overall safety protocols necessary for operation. \\ 
\rowcolor[HTML]{FFFFFF} 
Environmental & The environmental impact is addressed with a focus on GHG, PM, NOx emissions and noise levels. \\ 
\rowcolor[HTML]{FFFFFF} 
Economic & Includes capital expenditures, fixed operating and maintenance costs, start-up and fuel costs. \\ \hline
\end{tabular}
\caption{Key dimensions for qualitative technology assessment.}
\label{tab:dimensions_qualitative_assessment}
\end{table}

Table \ref{fig:technologies_selection} provides an overview of the variety of technologies that were assessed as well as a schematic of the decision matrix and the relative performance of each technology. The color of each cell represents how well a specific technology performs in a particular area. The suitability of EDGs was assessed by examining their previously outlined strengths and weaknesses, serving as the base case for comparison. Dark ($\checkmark$) and light (-) green shades indicate high and adequate suitability while yellow shades (x) indicate low suitability. The distribution of colors highlights strengths and limitations for each type of technology and also distinguishes variations within individual categories. 

It can be seen that conventional EDGs and natural gas generators perform well across a wide range of criteria but fall short on the environmental aspect. All other technologies offer advantages in terms of environmental impact but have other limitations. For example, some clean fuels face safety challenges linked to leakage or explosiveness. Batteries are limited by their high upfront investment costs and low volumetric energy density. This table, combined with expert judgment and a thorough review of literature, forms the basis for the selection of a subset of alternatives. The technologies chosen for further assessment include conventional EDGs as the base case, clean fuels (biofuel \& ammonia generators, hydrogen fuel cells, direct methanol fuel cells), rechargeable batteries (lithium-ion, iron-air), primary batteries (aluminum-air), and onsite VRE (solar PV).

\begin{table}[]
\centering
\footnotesize 
\setlength{\tabcolsep}{4pt} 
\renewcommand{\arraystretch}{1} 
\begin{tabular}{lcccccc}
\rowcolor[HTML]{E8E8E8} 
\diagbox{\textbf{Technology}}{\textbf{Factor}}        & \textbf{Technical}       & \textbf{Physical}        & \textbf{Maintenance}    & \textbf{Safety}          & \textbf{Environment}   & \textbf{Economic}        \\
Diesel Generator             & \cellcolor[HTML]{B1D68B} - & \cellcolor[HTML]{63BE7B} $\checkmark$ & \cellcolor[HTML]{B1D68B} - & \cellcolor[HTML]{63BE7B} $\checkmark$ & \cellcolor[HTML]{FFEF9C} x & \cellcolor[HTML]{63BE7B} $\checkmark$ \\
Natural Gas Generator        & \cellcolor[HTML]{B1D68B} - & \cellcolor[HTML]{B1D68B} - & \cellcolor[HTML]{B1D68B} - & \cellcolor[HTML]{B1D68B} - & \cellcolor[HTML]{FFEF9C} x & \cellcolor[HTML]{63BE7B} $\checkmark$ \\
\rowcolor[HTML]{E8E8E8} 
\textbf{Clean Fuels}         &                          &                          &                          &                          &                          &                          \\
Ammonia                      & \cellcolor[HTML]{B1D68B} - & \cellcolor[HTML]{63BE7B} $\checkmark$ & \cellcolor[HTML]{B1D68B} - & \cellcolor[HTML]{FFEF9C} x & \cellcolor[HTML]{63BE7B} $\checkmark$ & \cellcolor[HTML]{B1D68B} - \\
Methanol                     & \cellcolor[HTML]{B1D68B} - & \cellcolor[HTML]{63BE7B} $\checkmark$ & \cellcolor[HTML]{FFEF9C} x & \cellcolor[HTML]{FFEF9C} x & \cellcolor[HTML]{63BE7B} $\checkmark$ & \cellcolor[HTML]{B1D68B} - \\
Synthetic Diesel             & \cellcolor[HTML]{B1D68B} - & \cellcolor[HTML]{63BE7B} $\checkmark$ & \cellcolor[HTML]{B1D68B} - & \cellcolor[HTML]{63BE7B} $\checkmark$ & \cellcolor[HTML]{B1D68B} - & \cellcolor[HTML]{FFEF9C} x \\
Biofuel                      & \cellcolor[HTML]{B1D68B} - & \cellcolor[HTML]{63BE7B} $\checkmark$ & \cellcolor[HTML]{FFEF9C} x & \cellcolor[HTML]{63BE7B} $\checkmark$ & \cellcolor[HTML]{B1D68B} - & \cellcolor[HTML]{B1D68B} - \\
Synthetic Methane            & \cellcolor[HTML]{FFEF9C} x & \cellcolor[HTML]{FFEF9C} x & \cellcolor[HTML]{B1D68B} - & \cellcolor[HTML]{FFEF9C} x & \cellcolor[HTML]{B1D68B} - & \cellcolor[HTML]{B1D68B} - \\
Hydrogen                     & \cellcolor[HTML]{FFEF9C} x & \cellcolor[HTML]{B1D68B} - & \cellcolor[HTML]{63BE7B} $\checkmark$ & \cellcolor[HTML]{FFEF9C} x & \cellcolor[HTML]{63BE7B} $\checkmark$ & \cellcolor[HTML]{B1D68B} - \\
\rowcolor[HTML]{E8E8E8} 
\textbf{Secondary Batteries} &                          &                          &                          &                          &                          &                          \\
Lithium-Ion                  & \cellcolor[HTML]{63BE7B} $\checkmark$ & \cellcolor[HTML]{FFEF9C} x & \cellcolor[HTML]{63BE7B} $\checkmark$ & \cellcolor[HTML]{B1D68B} - & \cellcolor[HTML]{B1D68B} - & \cellcolor[HTML]{FFEF9C} x \\
Lithium-Iron Phosphate       & \cellcolor[HTML]{B1D68B} - & \cellcolor[HTML]{FFEF9C} x & \cellcolor[HTML]{63BE7B} $\checkmark$ & \cellcolor[HTML]{63BE7B} $\checkmark$ & \cellcolor[HTML]{63BE7B} $\checkmark$ & \cellcolor[HTML]{FFEF9C} x \\
Lithium-Manganese Oxide      & \cellcolor[HTML]{B1D68B} - & \cellcolor[HTML]{FFEF9C} x & \cellcolor[HTML]{63BE7B} $\checkmark$ & \cellcolor[HTML]{B1D68B} - & \cellcolor[HTML]{63BE7B} $\checkmark$ & \cellcolor[HTML]{FFEF9C} x \\
Lead-Acid                    & \cellcolor[HTML]{FFEF9C} x & \cellcolor[HTML]{FFEF9C} x & \cellcolor[HTML]{63BE7B} $\checkmark$ & \cellcolor[HTML]{FFEF9C} x & \cellcolor[HTML]{FFEF9C} x & \cellcolor[HTML]{B1D68B} - \\
Vanadium Redox               & \cellcolor[HTML]{B1D68B} - & \cellcolor[HTML]{FFEF9C} x & \cellcolor[HTML]{63BE7B} $\checkmark$ & \cellcolor[HTML]{FFEF9C} x & \cellcolor[HTML]{63BE7B} $\checkmark$ & \cellcolor[HTML]{B1D68B} - \\
Zinc-Air                     & \cellcolor[HTML]{FFEF9C} x & \cellcolor[HTML]{B1D68B} - & \cellcolor[HTML]{B1D68B} - & \cellcolor[HTML]{B1D68B} - & \cellcolor[HTML]{63BE7B} $\checkmark$ & \cellcolor[HTML]{B1D68B} - \\
Magnesium-Air                & \cellcolor[HTML]{FFEF9C} x & \cellcolor[HTML]{B1D68B} - & \cellcolor[HTML]{B1D68B} - & \cellcolor[HTML]{B1D68B} - & \cellcolor[HTML]{63BE7B} $\checkmark$ & \cellcolor[HTML]{B1D68B} - \\
Iron-Air                     & \cellcolor[HTML]{63BE7B} $\checkmark$ & \cellcolor[HTML]{FFEF9C} x & \cellcolor[HTML]{63BE7B} $\checkmark$ & \cellcolor[HTML]{B1D68B} - & \cellcolor[HTML]{63BE7B} $\checkmark$ & \cellcolor[HTML]{B1D68B} - \\
Molten Salt                  & \cellcolor[HTML]{B1D68B} - & \cellcolor[HTML]{FFEF9C} x & \cellcolor[HTML]{FFEF9C} x & \cellcolor[HTML]{63BE7B} $\checkmark$ & \cellcolor[HTML]{63BE7B} $\checkmark$ & \cellcolor[HTML]{FFEF9C} x \\
\rowcolor[HTML]{E8E8E8} 
\textbf{Primary Batteries}   &                          &                          &                          &                          &                          &                          \\
Alkaline Manganese           & \cellcolor[HTML]{FFEF9C} x & \cellcolor[HTML]{FFEF9C} x & \cellcolor[HTML]{B1D68B} - & \cellcolor[HTML]{B1D68B} - & \cellcolor[HTML]{B1D68B} - & \cellcolor[HTML]{FFEF9C} x \\
Zinc-Carbon                  & \cellcolor[HTML]{FFEF9C} x & \cellcolor[HTML]{FFEF9C} x & \cellcolor[HTML]{B1D68B} - & \cellcolor[HTML]{FFEF9C} x & \cellcolor[HTML]{B1D68B} - & \cellcolor[HTML]{B1D68B} - \\
Aluminum-Air                & \cellcolor[HTML]{FFEF9C} x & \cellcolor[HTML]{B1D68B} - & \cellcolor[HTML]{B1D68B} - & \cellcolor[HTML]{B1D68B} -  & \cellcolor[HTML]{B1D68B} - & \cellcolor[HTML]{B1D68B} - \\
Silver Oxide                 & \cellcolor[HTML]{FFEF9C} x & \cellcolor[HTML]{FFEF9C} x & \cellcolor[HTML]{FFEF9C} x & \cellcolor[HTML]{B1D68B} - & \cellcolor[HTML]{B1D68B} - & \cellcolor[HTML]{FFEF9C} x \\
Lithium-Carbon Fluide        & \cellcolor[HTML]{FFEF9C} x & \cellcolor[HTML]{B1D68B} - & \cellcolor[HTML]{B1D68B} - & \cellcolor[HTML]{FFEF9C} x & \cellcolor[HTML]{B1D68B} - & \cellcolor[HTML]{FFEF9C} x \\
\rowcolor[HTML]{E8E8E8} 
\textbf{Onsite VRE}          &                          &                          &                          &                          &                          &                          \\
Wind                         & \cellcolor[HTML]{63BE7B} $\checkmark$ & \cellcolor[HTML]{B1D68B} - & \cellcolor[HTML]{63BE7B} $\checkmark$ & \cellcolor[HTML]{63BE7B} $\checkmark$ & \cellcolor[HTML]{63BE7B} $\checkmark$ & \cellcolor[HTML]{B1D68B} - \\
Solar PV                     & \cellcolor[HTML]{63BE7B} $\checkmark$ & \cellcolor[HTML]{B1D68B} - & \cellcolor[HTML]{63BE7B} $\checkmark$ & \cellcolor[HTML]{63BE7B} $\checkmark$  & \cellcolor[HTML]{63BE7B} $\checkmark$ & \cellcolor[HTML]{63BE7B} $\checkmark$ \\
Solar Thermal                & \cellcolor[HTML]{B1D68B} - & \cellcolor[HTML]{B1D68B} - & \cellcolor[HTML]{B1D68B} - & \cellcolor[HTML]{B1D68B} - & \cellcolor[HTML]{63BE7B} $\checkmark$ & \cellcolor[HTML]{FFEF9C} x \\
\rowcolor[HTML]{E8E8E8} 
\textbf{Thermal}             &                          &                          &                          &                          &                          &                          \\
Molten Salt                  & \cellcolor[HTML]{FFEF9C} x & \cellcolor[HTML]{FFEF9C} x & \cellcolor[HTML]{B1D68B} - & \cellcolor[HTML]{63BE7B} $\checkmark$ & \cellcolor[HTML]{63BE7B} $\checkmark$ & \cellcolor[HTML]{FFEF9C} x \\
Crushed Rock                 & \cellcolor[HTML]{FFEF9C} x & \cellcolor[HTML]{FFEF9C} x & \cellcolor[HTML]{63BE7B} $\checkmark$ & \cellcolor[HTML]{63BE7B} $\checkmark$ & \cellcolor[HTML]{63BE7B} $\checkmark$ & \cellcolor[HTML]{FFEF9C} x
\end{tabular}
\caption{Overview of alternative technologies and qualitative assessment.}
\label{fig:technologies_selection}
\end{table}

\subsection{BPS characterization}

This study considers BPSs as standalone systems and assumes no connection to distribution or transmission lines during power outages. Interconnections with other off-site BPSs are not allowed. A life-cycle approach for emissions accounting is used, including emissions from both onsite fuel combustion and upstream fuel production. This framework is often referred to as scope 2 emission accounting. For secondary batteries, the carbon footprint of the electricity used for charging is included. Primary batteries are considered analogous to fuel sources due to their single-use nature. Their carbon footprint thus also evaluates the production emissions. The assessment excludes embodied emissions from the manufacturing of fuel cells, generators, and secondary batteries. Fuel degradation necessitates periodic fuel replacement. Combined with regular testing of the BPS this leads to additional costs and emissions. The analysis assumes an optimized maintenance schedule that aligns system testing with fuel replacement intervals. The fuel consumption of the replacement process is included in the model as an upfront cost and the emissions generated from burning the entire fuel stored are factored in. It is assumed that the cost of fuel polishing is equal to the cost of fuel replacement. \cite{fuel_polishing} 

BPSs are designed to autonomously supply power for extended periods of time and many operators do not consider fuel purchases during power outages, hereafter referred to as \textit{emergency fuel purchases}. For this reason, the base case scenarios presented in this report assume standalone BPSs with no external emergency fuel purchases during outages. However, evidence shows that while extreme events may cause disruptions, fuel supply chains typically retain some functionality, enabling emergency fuel delivery.\cite{yang2022optimizing} Therefore, this study incorporates a number of distinct scenarios with the option of emergency fuel purchases to reflect these circumstances. Fuel prices can increase significantly during exteme events, as observed in the October 2024 Florida hurricane. To model this, emergency fuel purchases are assumed to be more expensive than under normal conditions. Emergency fuel purchases can include any clean fuel or primary batteries. Charged secondary batteries are excluded from emergency fuel purchases. The techno-economic assessment of battery energy storage systems (BESSs) is done in accordance with the 2024 annual technology baseline (ATB) for a 4-hour nameplate utility-scale BESS.

\subsection{Power outage analysis}

To avoid the need to optimize BPS design operation using all historical outages, we developed a methodology to create representative outages. The methodology combines pattern identification, correlation analysis, and clustering techniques. As a first step, we analyzed regional differences and patterns in power outages using a comprehensive dataset compiled by DOE researchers. \cite{brelsford2024dataset} The dataset offers a detailed view of electricity outages across the US from 2014 to 2023, covering up to 92\% of customers with 15-minute resolution. Outages shorter than 15 minutes may not be captured by the dataset due to the return period of the web parser system.\cite{brelsford2024dataset} We extracted data on outages in the US state of Massachusetts between 2020 and 2023. This came to 55,000 outages across 14 counties. We used this data to estimate the overall outage probability for customers. It is assumed that the dataset accurately represents the entire customer population. This assumption is considered valid because the dataset consistently covers over 89\% of customers in Massachusetts throughout the four-year period. \cite{brelsford2024dataset} The total outage probability is calculated as follows:

\begin{equation}
\label{eq:outage_prob}
\mathbb{P}_{total} = \frac{\sum_{j=1}^{J}\sum_{i=1}^{N_j} C_{i,j} \cdot D_{i,j}}{M \cdot \sum_{j=1}^{J} T_j}
\end{equation}

$C_{i,j}$ represents the number of customers affected by the i-th outage in county j, $D_{i,j}$ stands for the duration of the i-th outage in county j measured in minutes, $N_j$ is the number of outages in county j, M contains the number of minutes within the time-horizon and $T_j$ is the total number of customers in county j. 

In the second step, a k-means clustering approach is used to partition power outage data into representative groups. K-means clustering works by iteratively assigning each data point to the cluster with the nearest centroid and minimizing the overall distance between individual data points and their respective cluster centers. A given dataset $X = {x_i \mid i = 1, 2, ..., n}$ of size n with d-dimensional data is clustered into k clusters $C = {c_j \mid j = 1, 2, ..., k}$. \cite{ikotun2023k} This is achieved by minimizing the sum of the squared error within each cluster:

\begin{equation}
\text{min} \hspace{0.5cm} J(C) = \sum_{k=1}^{K} \sum_{x_i \in c_k} ||x_i - \mu_k||^2
\end{equation}

While finding the optimal cluster arrangement is an NP-hard problem, the k-means clustering algorithm is fast and known to produce good results in a wide range of applications. \cite{na2010research} Despite its popularity, the method has some limitations. Choosing the wrong number of clusters can lead to suboptimal results \cite{yuan2019research} \cite{chong2021k}. Furthermore, noise and outliers can distort the clustering results. \cite{gan2017k} These limitations have to be considered when designing the clustering approach.

\subsection{Stochastic optimization}

We frame the scenario-based stochastic optimization approach from the perspective of an entity investing in a self-contained standby power system at a single location. The model uses a static approach and determines the optimal BPS design for a fixed point in the future, rather than continuously adapting to changing conditions over time. This static approach is widely used in expansion planning problems. \cite{ruiz2015robust} To account for the inherent uncertainty of power outages, the model incorporates the variability of outage duration and power demand using different scenarios, each weighted by a probability of occurrence. A comprehensive approach would involve simulating multiple years with potentially numerous outages per year. Based on supporting literature \cite{outages_massachusetts} and DOE data \cite{brelsford2024dataset}, the study assumes a maximum of one outage per year. This justifies the assumption of independence of individual outages and allows for the temporal decoupling of separate outage events. This approach reduces the computational burden while preserving the key characteristics of power outage variability. A finite number of scenarios is identified using the clustering method described previously. The scenarios are represented by a limited number of distinct representative periods and a probability of occurrence. The underlying stochastic optimization formulation represents a two-stage stochastic program. In a first stage, the investment decisions are made before the exact scenario is revealed. Thus, the chosen BPS must be capable of meeting the power demand in all scenarios. In a second stage and after observing the actual demand, the optimal dispatch decisions are made. The resulting expected value of the objective function is minimized, while the entire set of constraints has to be satisfied in every representative period. 

The number of time steps within each representative period are calculated based on the longest power outage observed in the data set comprised of 55,000 outages in Massachusetts.\cite{brelsford2024dataset} This corresponds to a duration of roughly 5 days and the total number of time steps per representative period is set to 480, each representing a 15-minute interval. This 15-minute time resolution is necessary to accurately capture the impact of short-duration power outages, as has been outlined before. For a maximum of 400 representative periods, each containing 480 time steps, the computational burden of this model is equal to a 22-year, one-zone modeling in hourly resolution. Using this temporal resolution for the representative periods, the weights for the individual time steps in the model can be computed as follows: \\

\begin{equation}
\omega_t = \mathbb{P}_{total} \times \frac{N_s}{N_{total}} \times \frac{1}{N_{timesteps}}, \quad \forall t \in \mathcal{S}
\end{equation}

$\omega_t$ defines the weight of each time step within a representative period and thus determines the expected value of the objective function. $P_{total}$ is the overall outage likelihood, $N_s$ represents the number of outages with duration $s$ and $N_{total}$ is the total number of outages within the dataset. It is assumed that for a sufficiently large number of samples the sample proportion $\frac{N_s}{N_{total}}$ converges to the true probability $\mathbb{P}(Duration = s | Outage = true)$. The weights apply to all time steps within the representative period corresponding to the scenario $\mathcal{S}$. The aforementioned specifications give rise to the following mathematical formulation of the stochastic program. For readability purposes, only the decision variables and constraints directly linked to standalone BPSs consisting of generation and storage units are listed. The notation is adapted from the GenX capacity expansion model, jointly developed by the MIT Energy Initiative and the Princeton University ZERO lab. \cite{genx_source}. \\

\textbf{Objective function:}
\begin{align*}
    \text{min} \quad
    &\sum_{y \in \mathcal{G}} \left((\pi^{INVEST}_{y} \times  \Omega_{y}) + (\pi^{FOM}_{y} \times  \Delta^{total}_{y})\right) +  \\\
    &\sum_{y \in \mathcal{G}} \left( (\pi^{INVEST,storage}_{y} \times    \Omega^{storage}_{y}) + (\pi^{FOM,storage}_{y} \times  \Delta^{total,storage}_{y})\right) +  \\\
    & \sum_{y \in \mathcal{G}} \sum_{t \in \mathcal{T}} \left( \omega_{t}\times \pi^{VOM}_{y} \times \Theta_{y,t}\right) + \sum_{y \in \mathcal{G} } \sum_{t \in \mathcal{T}} \left( \omega_{t}\times\pi^{FUEL}_{t,y} \times (\beta^{top_{up}}_{t,y} + \rho^{premium} \times \beta^{purchase}_{t,y}) \right) + \\\
    & \sum_{y \in \mathcal{G}} \left(\rho^{replace} \times \Delta^{total,storage}_y \times \pi^{FUEL}_{t,y} - \sum_{t \in \mathcal{T}} \omega_t \times \pi^{FUEL}_{t,y} \times \beta^{top_{up}}_{t,y}\right) + \\\
    &\sum_{y \in \mathcal{G}} \sum_{t \in \mathcal{T}}\left(\omega_{t} \times \pi^{START}_{y} \times \chi_{y,t}\right) +  \\\
    &\sum_{y \in \mathcal{V}} \left( (\pi^{INVEST, pv}_{y} \times \Omega^{pv}_{y})
    + (\pi^{FOM, pv}_{y} \times  \Delta^{total,pv}_{y})\right)
\end{align*}

The objective function minimizes the total system costs, including investment, fixed operation and maintenance, and variable operational costs. It accounts for generation, storage, and renewable energy investments as well as fuel costs, emergency fuel purchases and generator start-ups. \\

\textbf{Decision variables:}
\begin{align*}
    & \Omega_{y},\Delta^{total}_y, \Omega_{y}^{storage},\Delta^{total,storage}_y  \geq 0, \quad \forall y \in \mathcal{G} \\
    & \Omega^{pv}_{y}, \Delta^{total,pv}_{y} \geq 0, \quad \forall y \in \mathcal{V}\\
    & \beta^{level}_{t,y}, \beta^{purchase}_{t,y}, \beta^{top\_up}_{t,y}, \Theta_{y,t}, \chi_{y,t} \geq 0, \quad \forall t \in \mathcal{T}, y \in \mathcal{G} 
\end{align*}

$\Omega_{y}$ and $\Delta^{total}_{y}$ represent the newly added and total installed capacity of generation units, with the same being true for $\Omega_{y}^{storage}$ and $\Delta^{total,storage}_{y}$ as well as $\Omega^{pv}_{y}$ and $\Delta^{total,pv}_{y}$. The variables $\beta^{level}_{t,y}$, $\beta^{purchase}_{t,y}$, and $\beta^{top\_up}_{t,y}$ manage fuel levels, emergency purchases, and fuel top-ups. $\Theta_{y,t}$ and $\chi_{y,t}$ track generation and number of start-ups. \\

\textbf{BPS Constraints:}
\begin{align*}
    & \beta^{level}_{t,y} \leq \Delta^{total,storage}_y, \quad \forall t \in \mathcal{T}, y \in \mathcal{G} \\\
    & \beta^{top\_up}_{t,y} = \Delta^{total,storage}_y - \beta^{level}_{t,y}, \quad \forall t \in \mathcal{T}_{end}, y \in \mathcal{G} \\\
    & \beta^{level}_{t,y} = \Delta^{total,storage}_y, \quad \forall t \in \mathcal{T}_{start}, y \in \mathcal{G} \\\
    & \beta^{level}_{t,y} = \beta^{level}_{t-1,y} - \Theta_{y,t} + \beta^{purchase}_{t,y}, \quad \forall t \in \mathcal{T}_{int}, y \in \mathcal{G} \\\
    & \beta^{purchase}_{t,y} =  0, \quad \forall t \in \mathcal{T}_{non-purchase}, y \in \mathcal{G}\\\
    & \beta^{purchase}_{t,y} =  0, \quad \forall t \in \mathcal{T}, y \in \mathcal{G}_{non-purchase}\\\
    & \beta^{purchase}_{t,y} \leq \beta^{purchase}_{Max}, \quad \forall t \in \mathcal{T}_{purchase}, y \in \mathcal{G}
\end{align*}

The constraints ensure that the fuel level cannot exceed the storage capacity at any time and that the fuel top-up amount equals the difference between storage capacity and the fuel level at the end of a representative period. Furthermore, the fuel level must be at maximum capacity at the beginning of a representative period and depends on the previous level, fuel consumption, and emergency purchases. Emergency purchases are only allowed during specific times and are capped at a maximum value. This scenario-based stochastic program allows the subsequent assessment of cost-efficient BPSs in a variety of settings. The analysis includes variations of power demand and demand response options as well as decarbonization pathways for BPSs.

\section{Results}
\subsection{Power outage characterization}

\begin{figure}[!htb]
\centering
    \includegraphics[width=0.8 \textwidth]{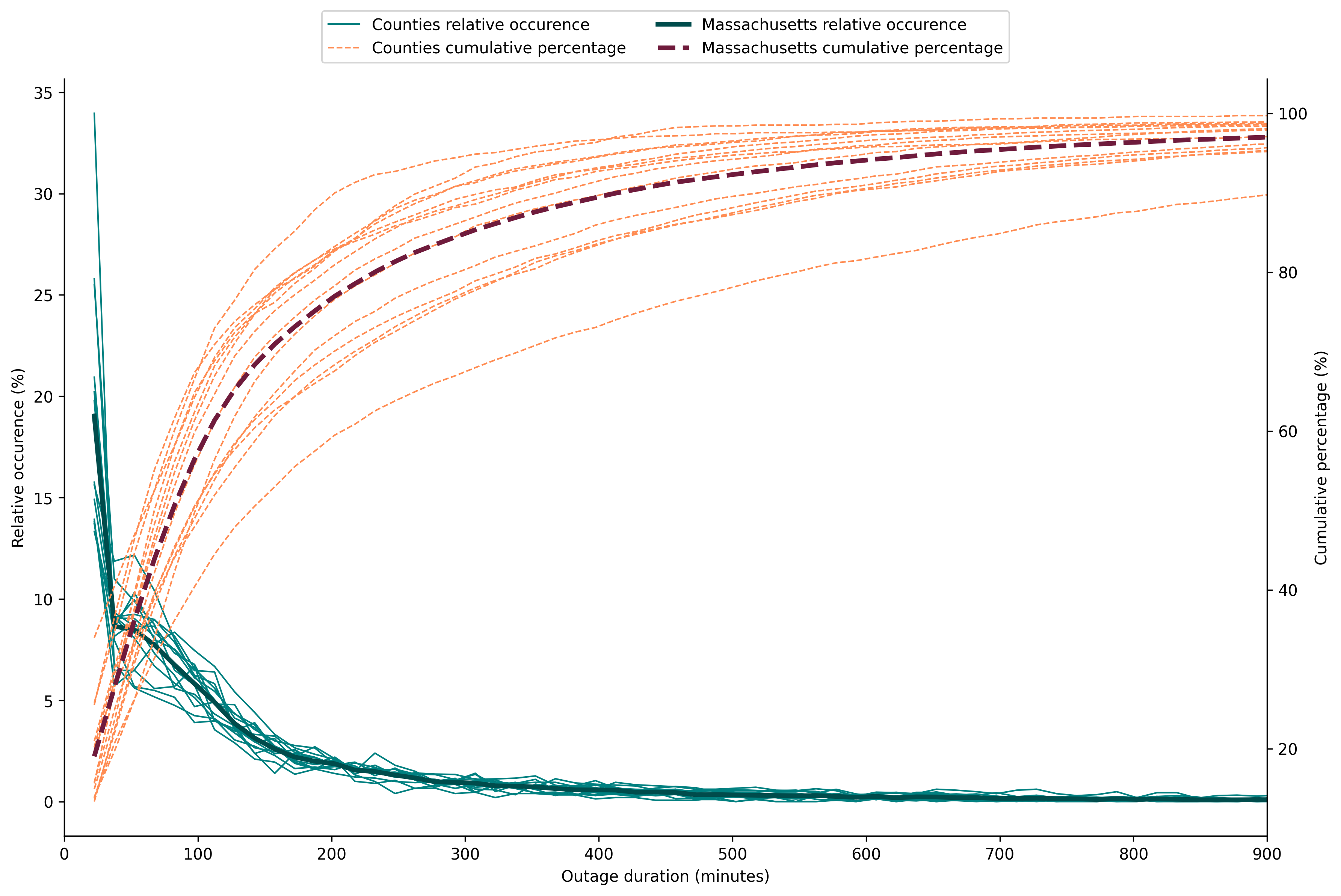}
    \caption{(left y-axis) The conditional probability of an outage lasting the specified duration if an outage occurs. The probability is given for each Massachusetts county and the state as a whole, between 2020-2023, based on the US department of energy DOE dataset. \cite{brelsford2024dataset}. (right y-axis) The cumulative conditional probability of an outage lasting at least the specified duration if an outage occurs. For visual clarity, the plot is limited to 900 minutes but the longest observed outage in the dataset lasted approximately 5 days.}
    \label{fig:power_outages_us}
\end{figure}

Figure~\ref{fig:power_outages_us} shows an extract of the 55,000 observed power outages in Massachusetts between 2020 and 2023, aggregated by county and ranked according to their duration. The figure shows that the distribution of outages across all counties follows a common pattern of frequent short-lasting and infrequent long-lasting outages. When aggregated to the state level, the data reveals that approximately 95\% of the 55,000 observed power outages last between 15 and 900 minutes (Massachusetts cumulative percentage). This can be explained by the fact that power outages are not only due to major storms but also due to smaller incidents such as fallen trees, motor vehicle accidents, equipment failure and wildlife interference. \cite{causes_power_outages}

\unskip

As a consequence of the observations in Figure~\ref{fig:power_outages_us}, the modeling granularity for the stochastic optimization is chosen to be 15 minutes. This distinguishes our approach from other studies which used temporal resolutions of an hour or more. Although longer outages are less likely they might have a considerable impact on overall system sizing and thus require accurate modeling. The dataset also reveals an overall outage probability of $P_{total} = 0.063\%$, aligning with findings from other research. \cite{outage_likelihood_overall_1} \cite{outage_likelihood_overall_2} 

As discussed previously, we sought to reduce the computational challenge of the stochastic optimization by identifying a set of representative outages. We find that power outages can be grouped based on several characteristics. The most important characteristics are duration, the month in which they occur, and time of the day at which they start. While outage duration is important for optimizing storage sizing, the starting point of an outage is essential to determining the load which the BPS must provide.

\begin{figure}[!htb]
    \centering
    \includegraphics[width= 0.8 \textwidth]{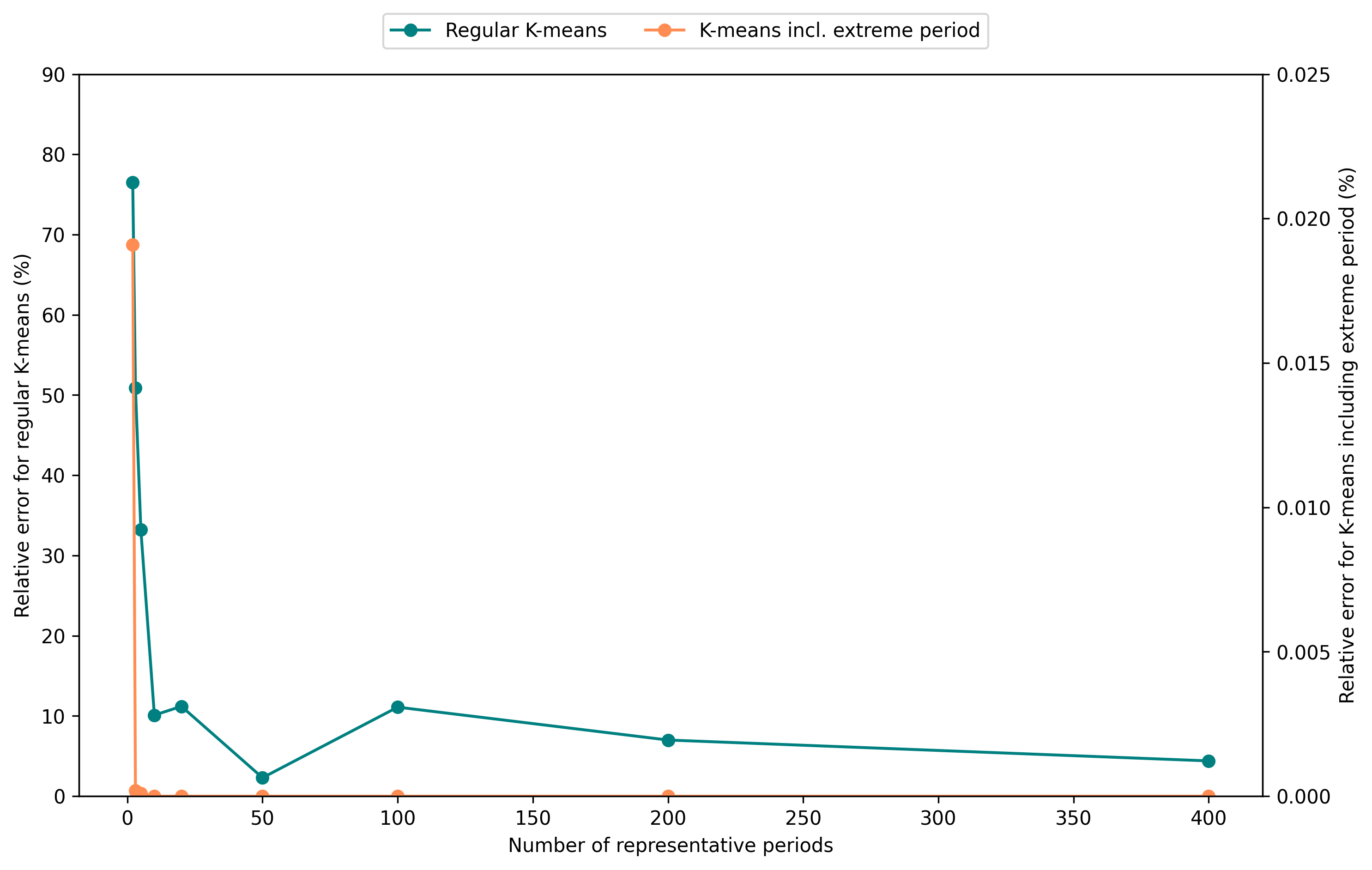}
    \caption{The absolute relative error of the objective function of the stochastic optimization using a given number of representative periods. The error is calculated as $|\text{Obj}_{\text{n rep}} - \text{Obj}_{g.t}| / \text{Obj}_{g.t}$, where $\text{Obj}_{\text{n rep}}$ is the objective function result of the stochastic optimization using $n$ representative periods and $\text{Obj}_{g.t.}$ is the objective function result using all of the outage data. The left y-axis shows the error for the case in which the representative periods are created through k-means clustering of the outage data. The right y-axis shows that same error but the representative periods also include the most extreme outage.}
    \label{fig:clustering_accuracy}
\end{figure}

Accuracy and computational complexity represent an important trade-off in high resolution energy system modeling. Figure~\ref{fig:clustering_accuracy} shows that using k-means clustering to find 10 representative periods leads to a 10\% error in the results. This error can be reduced to 4\% when running the model with 400 representative periods. Accuracy can be significantly further improved by including the most extreme outage period, characterized by peak power demand and maximum energy capacity, in the set of representative periods. Using this approach, an error of less than 0.01\% can be achieved with only 20 representative periods, drastically cutting the model's runtime from 2.5 hours (for a 400-cluster model) to 75 seconds. Peak power demand and maximum energy capacity are therefore crucial factors for an accurate cost assessment. For subsequent analysis, a set of 20 representative periods was used, including short-term and long-term power disruptions, the longest singular outage event recorded, and the outage event during which peak power demand was observed.

\subsection{Cost and emission drivers}
\begin{figure}[!htb]
    \centering
    \includegraphics[width=\textwidth]{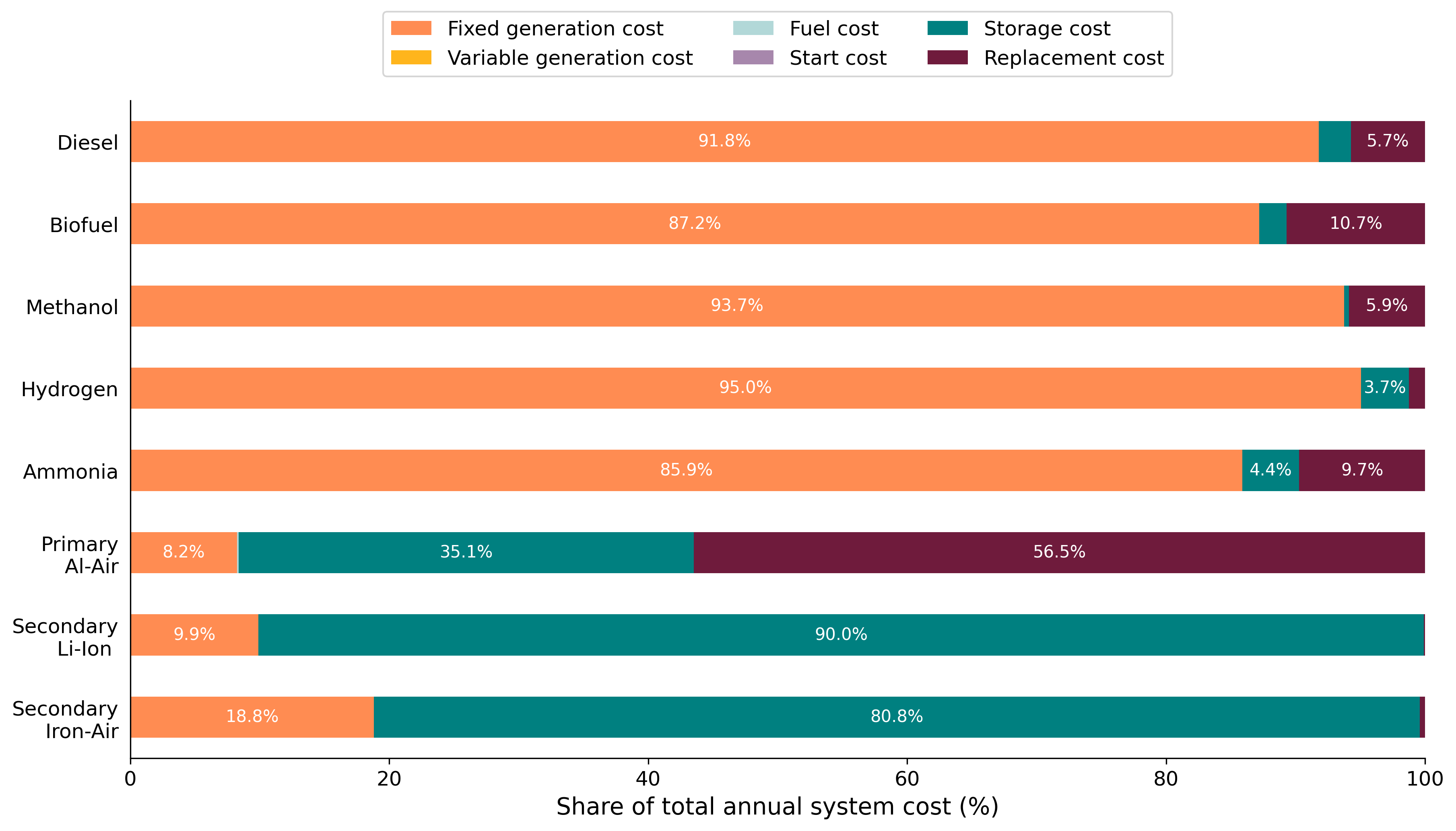}
    \caption{Cost structure of different technologies for a 8-hour outage at peak demand and outage likelihood $P_{total}= 0.0634\%$.}
    \label{fig:cost_structure_8h}
\end{figure}

Given the diverse cost structures of different technologies, it is essential to analyze the relative contribution of fixed costs, variable costs, fuel costs, and others to total annual system costs. Figure~\ref{fig:cost_structure_8h} displays the cost components for the technologies under investigation for a 8-hour outage at peak demand covered by the respective single-technology backup system. The plot shows that fixed generation costs dominate for generators and fuel cells. Furthermore, primary aluminum-air batteries have a lower ratio of fixed costs but suffer from high replacement costs. Secondary batteries have high energy storage costs, with iron-air being cheaper than lithium-ion in absolute terms. In comparison to other cost factors, startup and fuel costs prove to be minor contributors to the overall cost. This is due to the low overall likelihood of power outages. This analysis shows the importance of investment and fuel replacement costs and therefore the need for a sensitivity analysis to assess the impact of potential uncertainties in these cost parameters on the overall system costs.

Furthermore, the sensitivity of the weights of the stochastic optimization to the overall outage likelihood $\mathbb{P}_{total}$ is assessed. The analysis confirms the expected linear relationship between total system cost and overall outage likelihood. Even considering significant variations in total outage likelihood, the total system cost shows minimal sensitivity, changing by less than 0.02\%. Therefore, the chosen approach is relatively robust to variations in overall outage likelihood and uncertainties in outage probabilities do not have a significant impact on cost estimates. This finding, together with the significance of peak power demand and maximum energy capacity, enhances the broader applicability of the study's results to other geographic regions.

\begin{figure}[!htb]
    \centering
    
    \includegraphics[width= \textwidth]{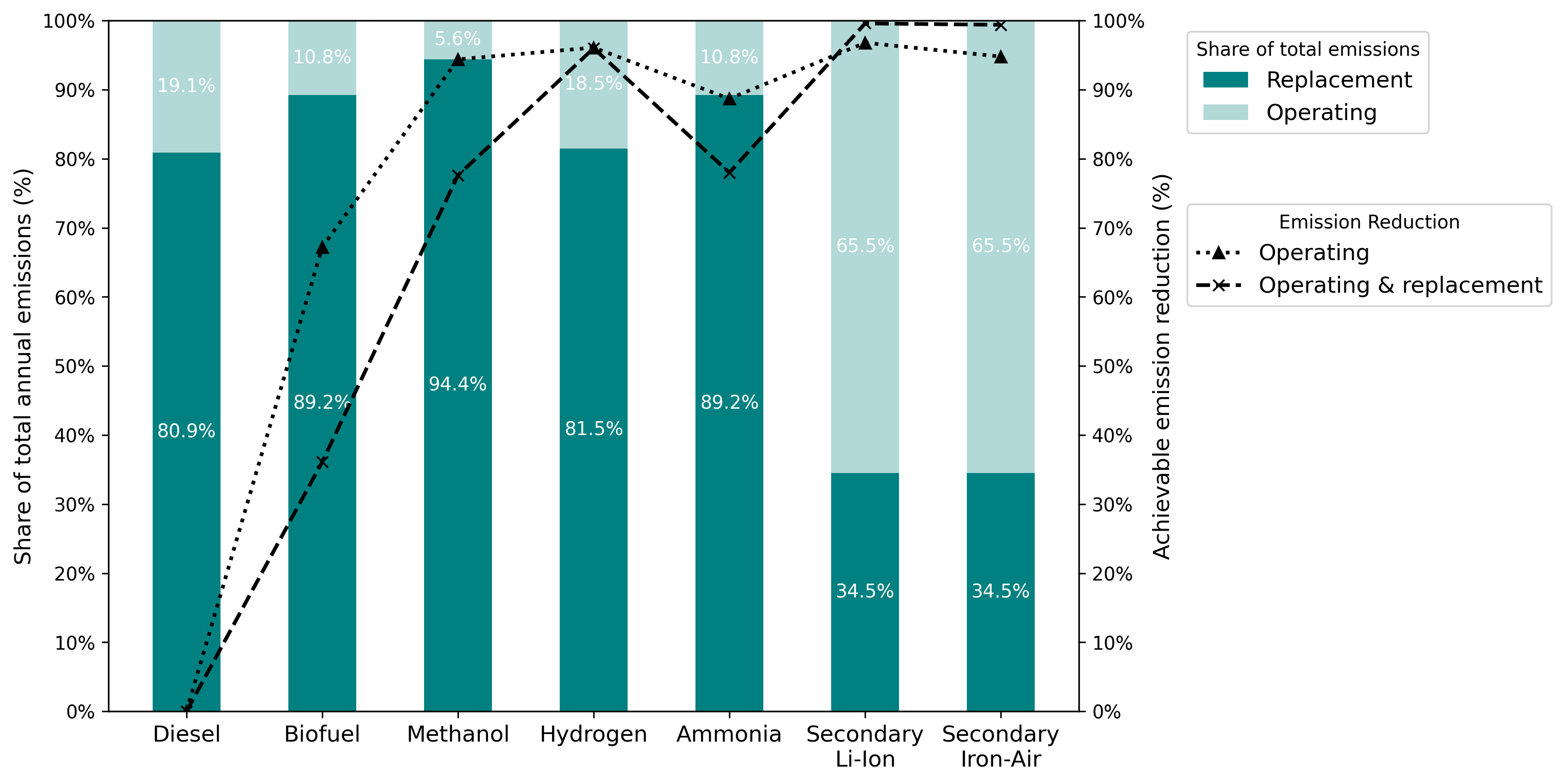}
    \caption{Emission comparison and reduction potential for different technologies for a 1-hour outage at peak demand and outage probability $P_{total}=0.0634\%$.}
    \label{fig:replacement_emissions}
\end{figure}

The analysis of key emission drivers of various technologies reveals that fuel replacement emissions constitute a substantial portion of total emissions. As can be seen in figure~\ref{fig:replacement_emissions} this impact is particularly significant for fuel-based technologies, where replacement emissions account for 80-95\% of total emissions, compared to approximately 35\% for battery technologies. Operating emissions constitute a small fraction of total emissions because of the infrequent power outage situations. The high significance of fuel replacement emissions has a direct impact on the emission reduction potential displayed on the secondary axis in figure~\ref{fig:replacement_emissions} and may lead to a bias in the perceived emission reduction potential.

Using the example of biofuel it can be seen that when accounting for operating emissions only, biofuel shows a 68\% emission reduction potential compared to diesel. However, accounting for fuel replacement emissions the net emission reduction potential decreases to 36\% due to the shorter shelf life of biofuel. This highlights the need for a comprehensive assessment of emission reduction potentials of BPSs especially for clean fuels, considering all relevant factors beyond operating emissions. An accurate emission accounting becomes especially relevant at longer outage durations with bigger onsite fuel storage and the relative importance of replacement emissions increases with the length of the longest outage that the system is sized for. 
Figure~\ref{fig:replacement_emissions} also illustrates the very high emission reduction potential of hydrogen and secondary batteries. It has to be noted that the omission of embodied emissions from this study may lead to a disproportional underestimation of the total emissions associated with the use of secondary batteries. As outlined before, this analysis assumes that users adhere to the recommended shelf-lives and replacement intervals of the various fuels and replace them at the time of system testing.

\subsection{Optimized sizing of hybrid BPSs}
\begin{figure}[H]
\centering
\subfloat[\centering \label{fig:battery_storage_role}]{%
    \includegraphics[width=7.7cm]{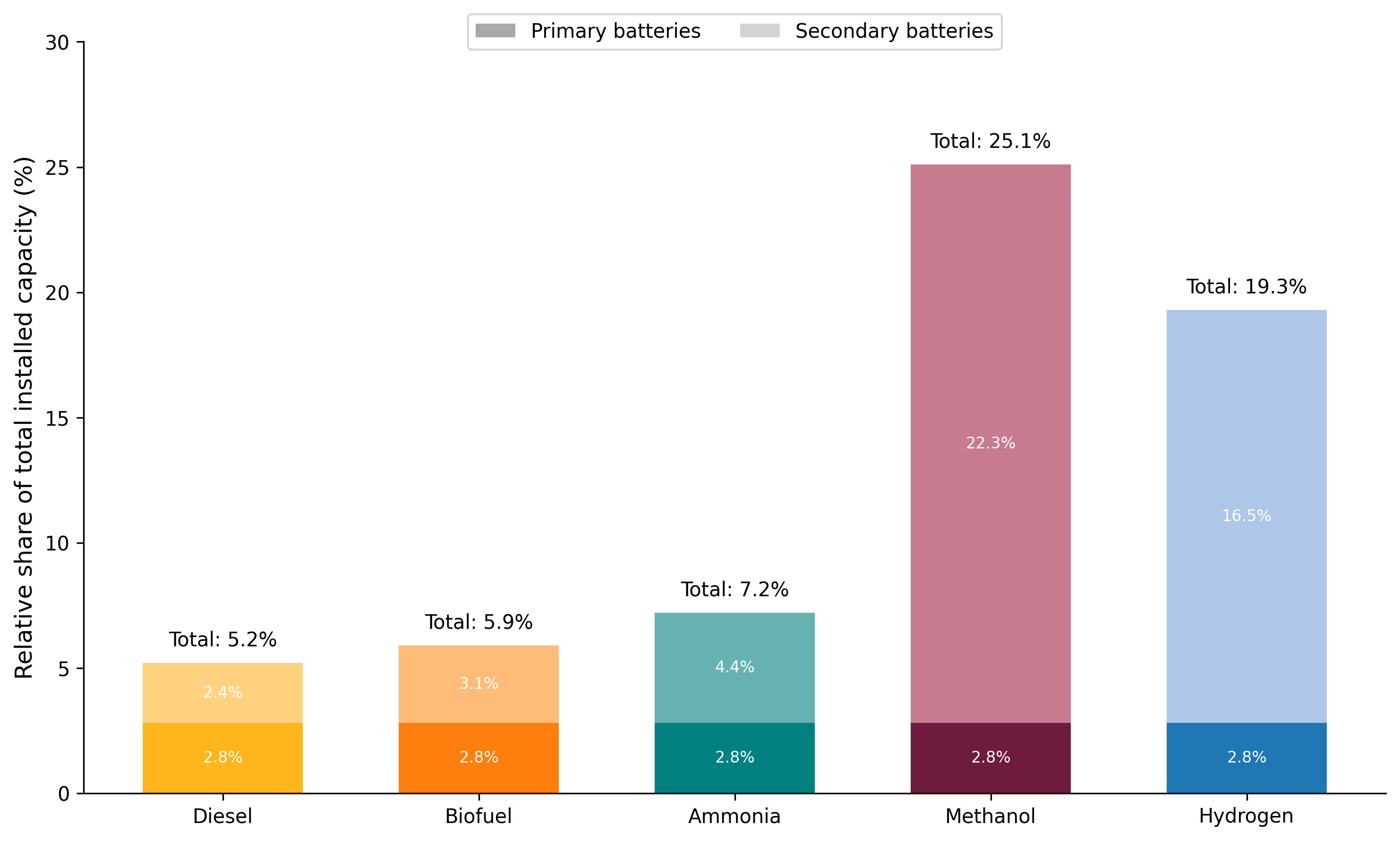}
}
\hfill
\subfloat[\centering \label{fig:time_evolution}]{%
    \includegraphics[width=7.7cm]{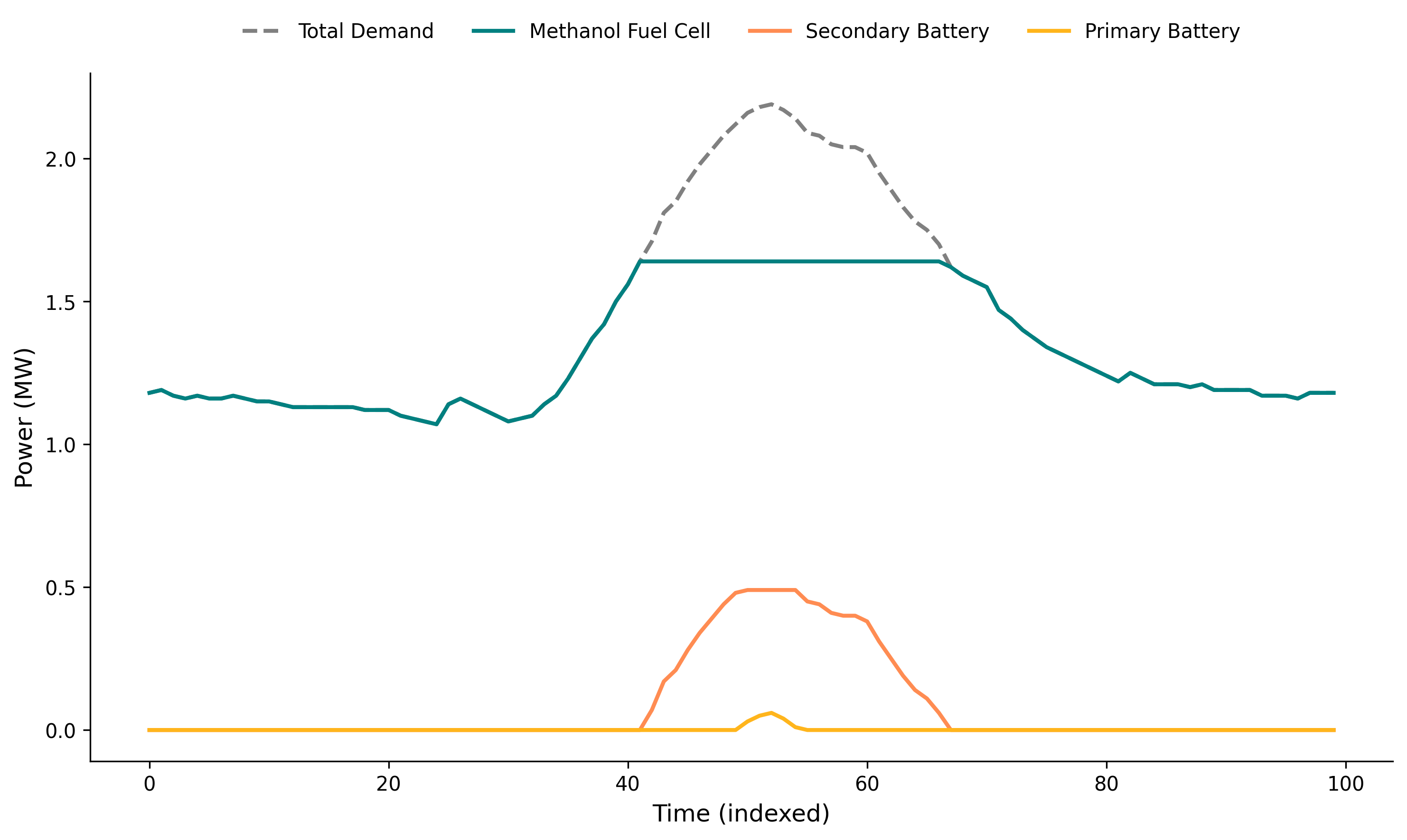}
}
\caption{Clean-fuel based hybrid BPSs. (\textbf{a}) Share of primary and secondary batteries in clean fuel-based hybrid systems for various base technologies. (\textbf{b}) Power contribution by technology during peak demand for a hybrid DMFC-battery system.}
\label{fig:flex_demand_figs}
\end{figure}

The following analysis examines the cost-effectiveness of hybrid BPSs and particularly focuses on the role of batteries in enhancing the economic viability of clean fuel-based systems. The results show that generators and fuel cells serve as core technologies in hybrid systems and that batteries can be used as a "cheap power, expensive energy" addition to clean fuel-based systems. While clean fuels offer cost-efficient long-term energy storage, batteries can lower total power costs by mitigating the need for clean fuels to be sized to meet peak demand. Figure~\ref{fig:battery_storage_role} shows the relative share of primary and secondary batteries in the overall generation mix of optimized hybrid BPSs. The graph must be interpreted as follows: It is optimal to size a biofuel generator to 94.1\% of peak power demand and cover the remaining 5.9\% with batteries.

The results show that the integration of primary and secondary batteries is cost-effective for each of the five clean fuel-based systems analyzed. While the share of primary batteries remains constant at 2.8\% across all base technologies, the share of secondary batteries varies depending on the specific clean fuel generator or fuel cell. This phenomenon is positively correlated with fixed generation costs. Clean fuels with higher power investment costs show higher shares of secondary batteries. For example, in DMFC systems, the optimal technology mix includes up to 22.3\% of secondary batteries. The observation of a constant share of primary batteries across all cases can be explained by the fact that primary batteries are solely sized for extreme peaks in power demand. A sample temporal generation profile of a hybrid BPS is illustrated in figure~\ref{fig:time_evolution}.

\begin{figure}[!htb]
    \centering
    
    \includegraphics[width= \textwidth]{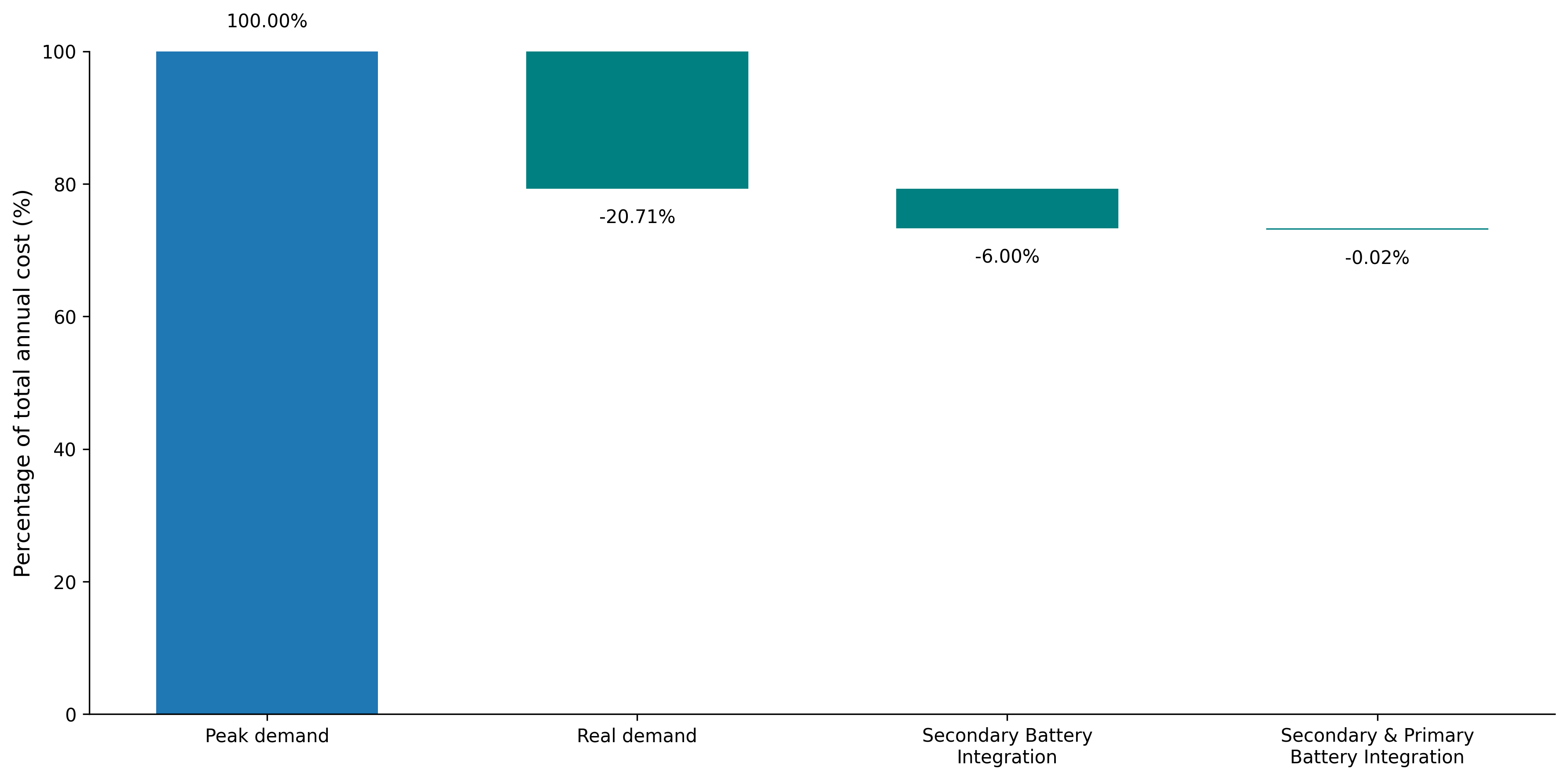}
    \caption{Analysis of cost savings for a direct methanol fuel cell (DMFC)-based system: Impact of optimized and demand-based system sizing and BESS integration.}
    \label{fig:true_demand_sizing}
\end{figure}

To further evaluate the cost-effectiveness of optimized systems, it is essential to examine the individual contributions of different cost-saving factors. In practice, the nameplate capacity of BPSs is typically sized within a range of 90\% to 125\% of peak demand \cite{genset_sizing} \cite{genset_sizing_2}. In the following analysis, it is assumed that nameplate capacity in this base case "peak sizing" is determined at exactly the value of peak load. The proposed model offers the capability to optimize backup systems based on a real demand pattern, which accounts for actual fluctuations in demand. Figure~\ref{fig:true_demand_sizing} shows how the costs of these different cases compare for a DMFC-based system. The peak and real demand cases exclude BESS integration with the fuel cell. The subsequent two cases explore this integration, first solely with secondary batteries and then using a combination of primary and secondary batteries.

The results show that significant cost reductions can be achieved with an optimized hybrid BPS sized to actual demand compared to peak demand. This cost reduction potential is threefold, namely reducing storage needs, decreasing fuel replacement costs and lowering power investment cost. Firstly, optimizing the system to accurately meet demand can significantly decrease the required storage capacity, leading to lower upfront investment costs. Secondly, the reduced storage capacity leads to a decrease in fuel replacement costs because a smaller amount of fuel has to be replaced on a regular basis. Lastly, by strategically integrating batteries, the resulting hybrid system can eliminate oversizing of expensive fuel cell or generator equipment by covering peaks in demand with secondary and primary batteries. Likewise, the total system cost can be reduced by up to 27\% compared to a DMFC-only system sized to peak demand.

\subsection{Optimal hybrid BPSs under emission limits}
\begin{figure}[!htb]
    \centering
    
    \includegraphics[width= \textwidth]{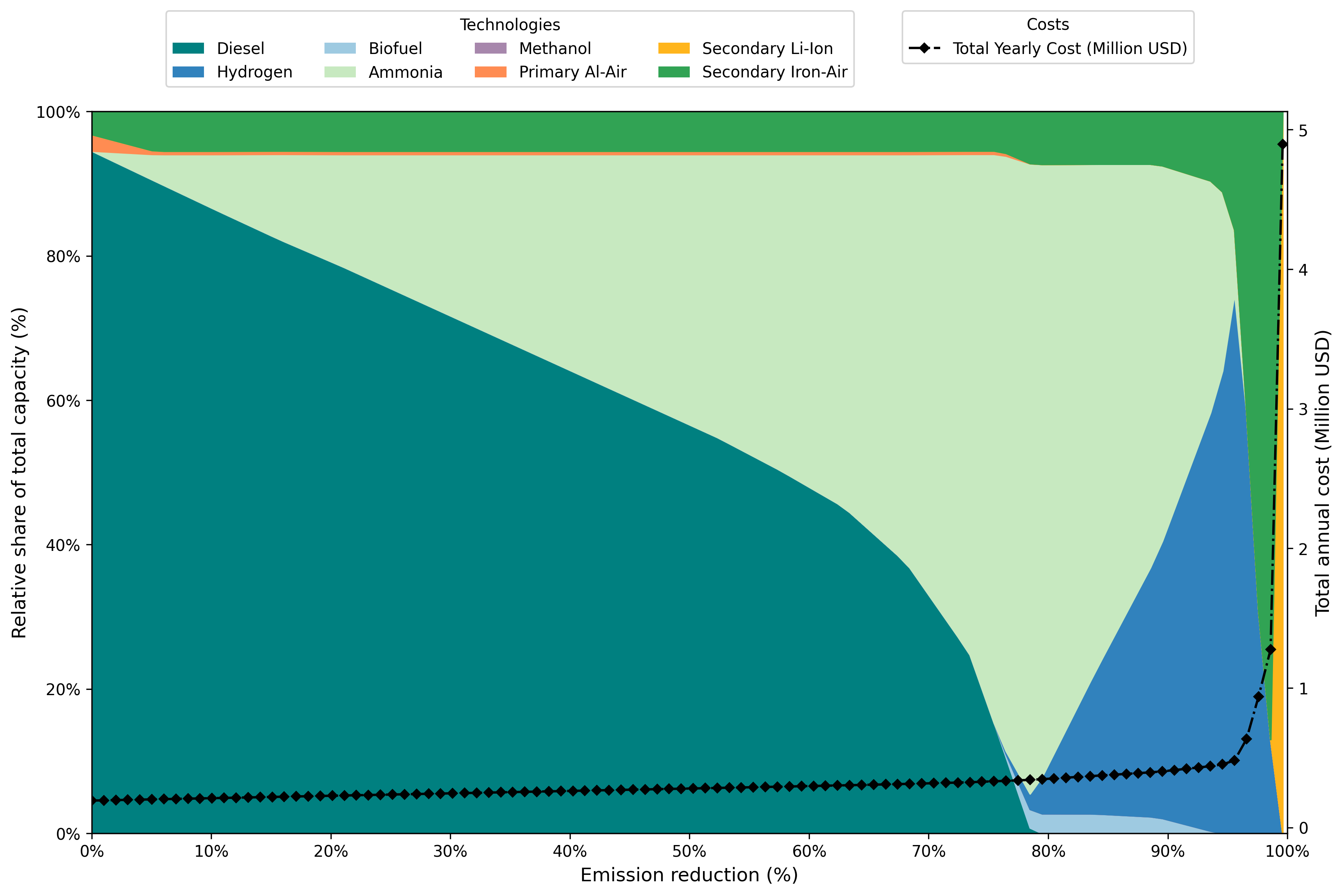}
    \caption{Optimal share of generating capacity and total annual system cost for varying CO\textsubscript{2} limits.}
    \label{fig:decarbonization_graph}
\end{figure}

Driven by stricter emission limits, the technological shifts and changes in total system cost are assessed. Figure~\ref{fig:decarbonization_graph} displays the share of generating capacity and the total annual system cost at different emission reduction targets. The analysis shows that a system combining fuel-based generation with battery storage is cost-effective throughout all decarbonization levels. Furthermore, as emission reduction targets become stricter, the share of EDGs declines and ammonia generators emerge as a valid alternative. A reduction of CO\textsubscript{2} emissions of around 79\% can be achieved by using hybrid BPSs composed of ammonia generators, sized to 93\% of peak demand, and secondary iron-air batteries, sized to 7\% of peak demand. The system essentially preserves the low capital expenditures (CAPEX) and high operational expenditures (OPEX) structure of EDGs and leverages the aforementioned complementarity of batteries and generators. 

At deep decarbonization levels the most cost-effective hybrid BPS is a combination of hydrogen fuel cells and secondary iron-air batteries. BPSs composed of hydrogen fuel cells and secondary iron-air batteries, with a capacity share of 81\% and 19\% respectively can achieve emission reductions of up to 96\% compared to EDGs. When focusing solely on scope 2 emissions, the theoretical limit of decarbonization can be reached with secondary batteries. The higher roundtrip efficiency of lithium-ion batteries enables them to outperform iron-air batteries in achieving higher emission reductions. Despite their theoretical possibility and occasional real-world implementation, these battery-only systems are not economically viable as outlined in the following paragraph.

Minimizing total system cost and emissions represent two conflicting objectives in optimization. The trade-off in this multi-objective optimization is visualized in the aforementioned figure~\ref{fig:decarbonization_graph}. As can be seen in the plot, reducing emissions comes with increased total annual system cost, highlighting these two competing objectives. The figure shows that significant emission reductions can be achieved at reasonable total system costs and that expenses rise sharply when pursuing the most ambitious decarbonization goals. For instance, emissions can be reduced by 42\% while increasing cost by 37\%. Higher decarbonization comes with a steep price tag, requiring already a 2.5-fold increase in total expenditures for an emission reduction of 95\%. Decarbonization levels of up to 75\% imply a CO\textsubscript{2} avoidance cost between 809 and 850 USD/ ton CO\textsubscript{2}. These costs remain higher than the estimated social cost of carbon, which ranges from 125 to 525 USD/ton CO\textsubscript{2}. \cite{tol2023social} It has to be noted that additional benefits such as reduced PM and NOx emissions, noise levels and reliability improvements are not quantified in this study.

Standalone secondary battery systems do not appear to be an economically viable option for backup systems. Compared to EDGs, standalone battery systems increase total annual system costs by a factor of 6 for iron-air and by a factor of 25 for li-ion systems. In particular, the high investment costs of li-ion batteries are not outweighed by their high cycling life and performance in the specific use case of emergency backup power.

\subsection{Optimal hybrid BPSs under emission limits with emergency fuel purchases}
\begin{figure}[!htb]
    \centering
    
    \includegraphics[width= \textwidth]{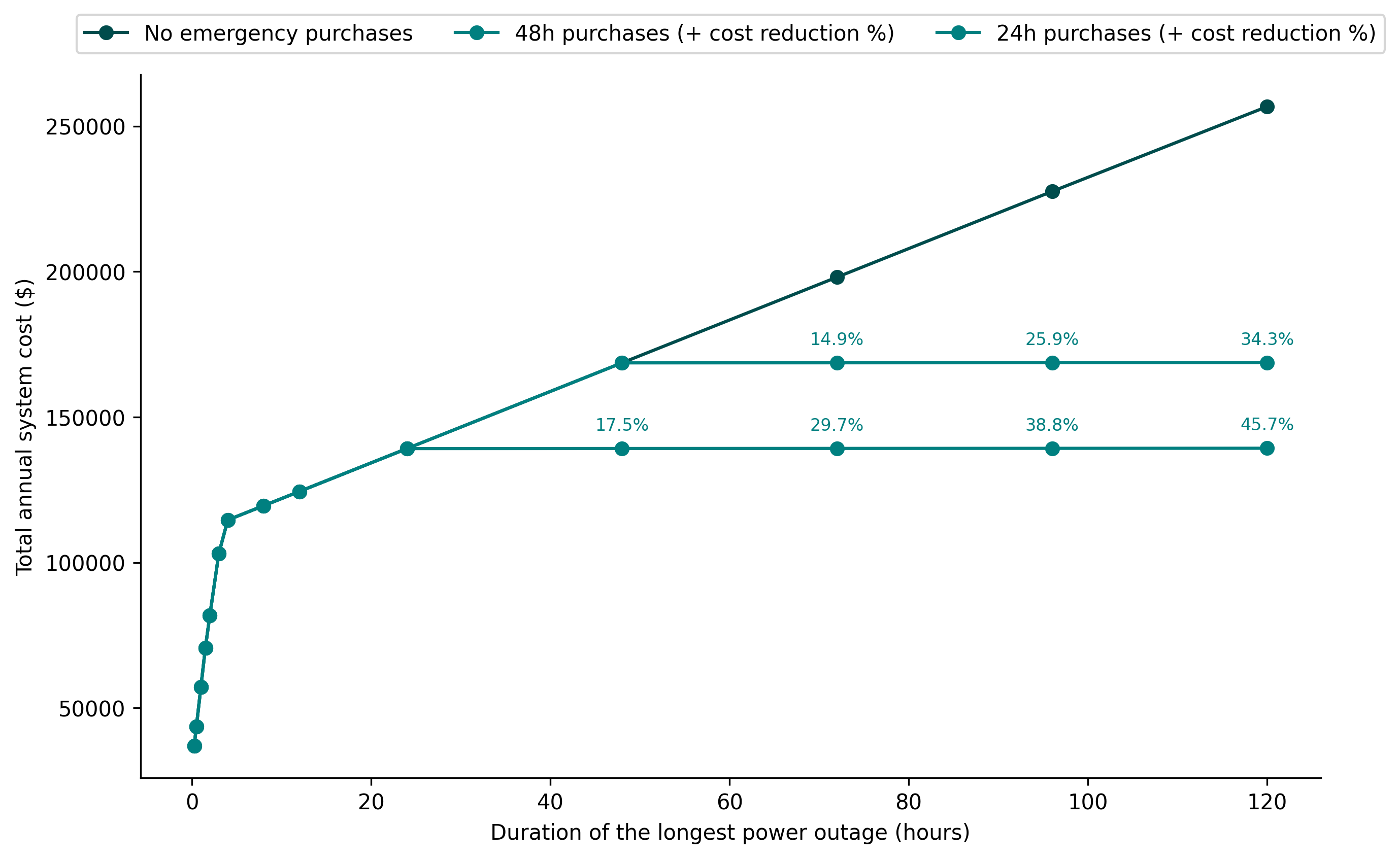}
    \caption{Potential cost savings of BPSs designed for various maximum outage durations and differing emergency purchase intervals.}
    \label{fig:emergency_purchases}
\end{figure}

Fuel supply chains typically retain some functionality even during extremely disruptive events. \cite{yang2022optimizing} Figure~\ref{fig:emergency_purchases} illustrates the relationship between the total annual system cost of a BPS and the maximum outage duration it is designed for. The figure also displays the impact of emergency fuel purchases on total annual system cost. When emergency fuel purchases are excluded, requiring all fuel to be stored onsite, the cost curve shows a steep increase due to the growing amount of energy that must be stored for longer outage durations. The change in slope around 5 hours marks the point where using clean-fuel based BPSs becomes more cost-effective than standalone BESSs which are more economical for short outages. The less steep slope indicates the more cost-effective long-term storage properties of clean fuel-based systems compared to standalone BESSs. When emergency fuel purchases are allowed the cost increase for systems designed for longer outages is mitigated. While the lines may look horizontal, their gradients remain slightly positive due to the price premium associated with emergency fuel purchases. 

The results show that the possibility for emergency fuel purchases can significantly reduce total annual system costs for BPSs that are designed for long outages. The majority of BPSs in use today are designed for long durations and thus fall into this category. The plot furthermore shows that shorter purchase intervals lead to greater cost reductions. In the case of a BPS designed to cover 5-day outages the cost reduction due to emergency fuel purchases amounts to 34\% and 46\% of total annual system cost for 48-hour and 24-hour purchase intervals respectively. This effect occurs because a smaller storage unit requires less upfront investment and reduces recurring fuel replacement costs.

The analysis considers a price premium of 43\% for emergency fuel purchases. A sensitivity study for this parameter shows that the changes in total annual system costs are robust towards changes of the exact value of this price premium. This observation can be explained by the low overall power outage likelihood.

\begin{figure}[!htb]
    \centering
    
    \includegraphics[width= \textwidth]{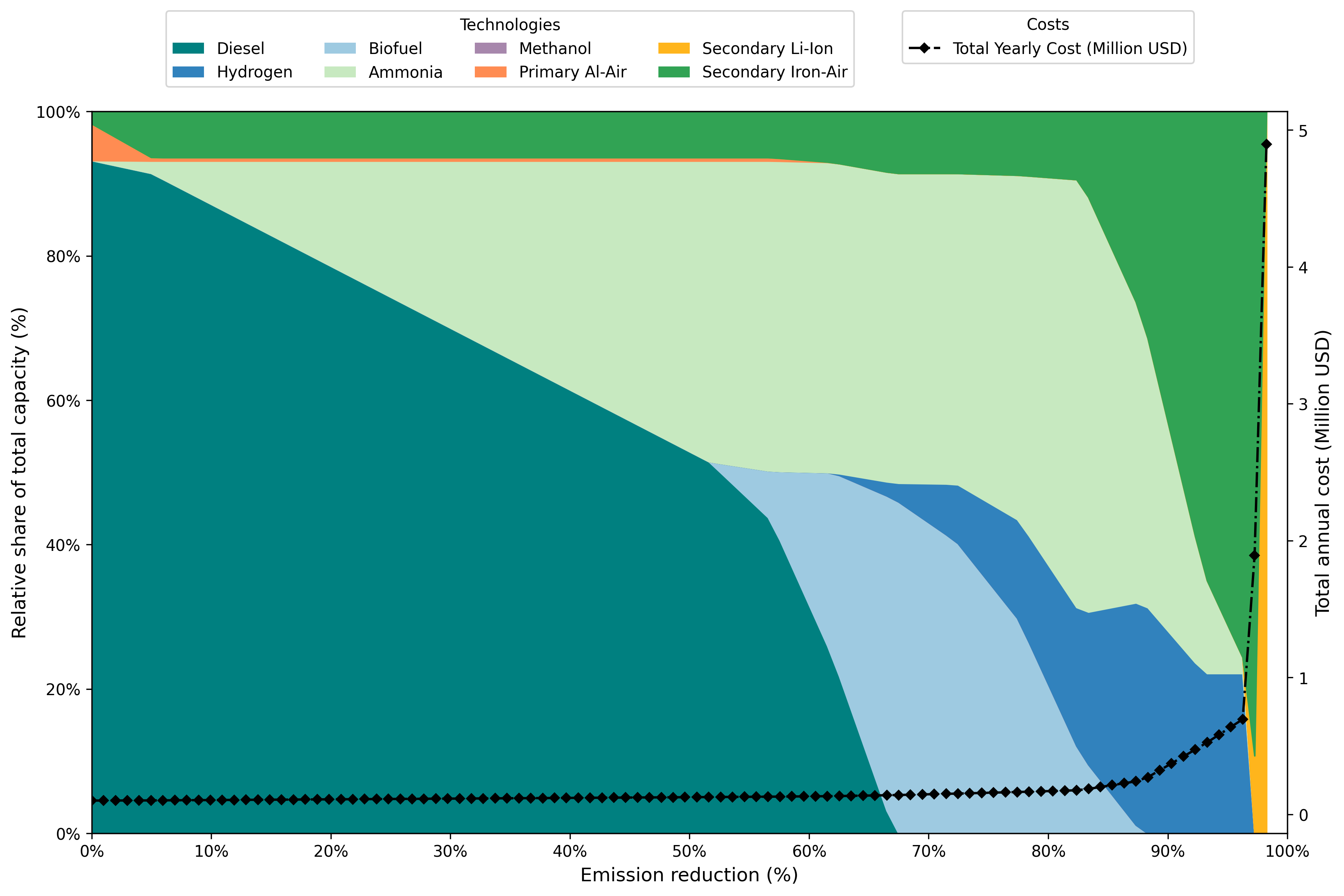}
    \caption{Optimal share of generating capacity and total annual system cost for varying CO\textsubscript{2} limits including volume-constrained 24-hour emergency purchases.}
    \label{fig:constrained_emergency_purchases}
\end{figure}

The impact of emergency fuel purchases on the optimal technology mix in decarbonized BPSs is negligible in the case of unconstrained fuel purchases and deliveries. However, the analysis is further extended to account for real-world constraints on fuel deliveries during power outages. Assuming volume-constrained deliveries and a standard truck capacity of 12'500 liters, the optimal technology mix is re-evaluated for different decarbonization levels. Figure~\ref{fig:constrained_emergency_purchases} shows a variety of decarbonization scenarios with volume-constrained 24-hour fuel deliveries. It has to be noted that the overall emissions of a system using emergency fuel purchases are lower compared to a regular system. The 0\% emission reduction case on the x-axis of the graph thus corresponds to the emissions of a hybrid diesel-battery BPS incorporating emergency fuel purchases. The results show that hybrid ammonia-secondary battery and hydrogen-secondary battery BPSs remain optimal across a wide range of decarbonization levels. Additionally, biofuel generators emerge as a cost-effective option. This is partially due to the higher volumetric energy density of biofuel compared to hydrogen, making it particularly advantageous for volume-constrained fuel deliveries.

\subsection{Sensitivity studies}

The objective of our study is to assess the cost-effectiveness and exact composition of decarbonized hybrid BPSs across a broad range of applications. Diverse backup power applications are characterized by different demand patterns as well as their potential for demand side response and the integration of onsite VRE resources. The results outline the potential of solar PV in the optimal generation mix of hybrid BPSs and quantify the impact of varying demand patterns and demand response options on optimal BPSs.

\subsubsection{Incorporating onsite solar PV}
\begin{figure}[!htb]
    \centering
    
    \includegraphics[width= \textwidth]{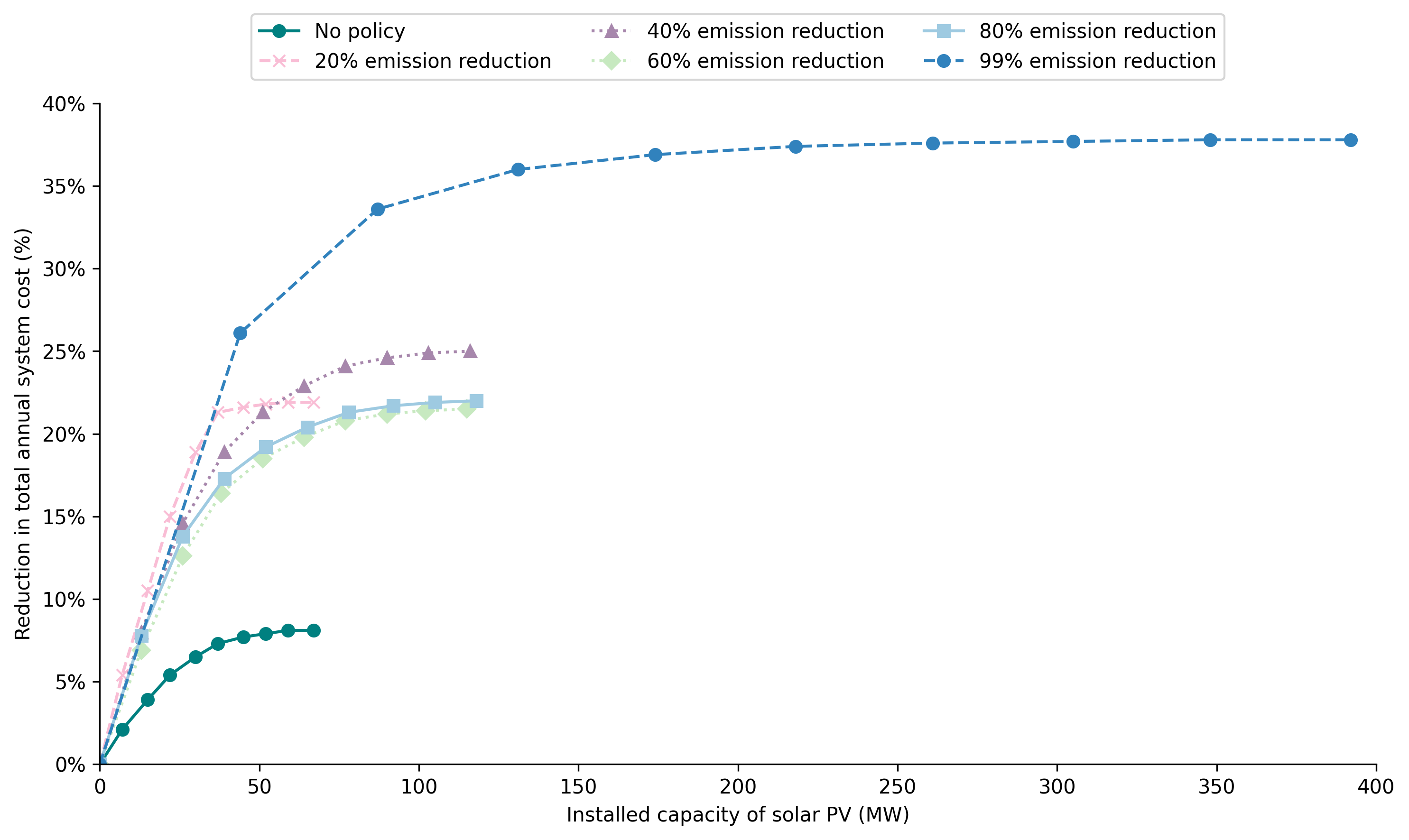}
    \caption{Cost reduction potential of solar PV at different decarbonization levels and unconstrained installed capacities using pro-rata cost accounting.}
    \label{fig:cost_reduction_solar_unbounded}
\end{figure}

The study investigates the potential of solar PV in the optimal generation mix of BPSs, taking into account the stochastic nature of solar power generation. The analysis is split into two parts and starts by assessing whether solar PV can be a cost-effective when solely used for backup power applications. The second part of the analysis investigates the potential of already installed solar PV. The results show that installing solar PV solely for standby power is not cost-effective even at low decarbonization levels. This finding aligns with the common usage of solar PV as continuous operating technology during non-outage times. To account for this primary use case, a pro-rata cost allocation is used as a second step of analysis to assess an economic "best case" for solar PV. The results show that using this pro-rata cost allocation and solely considering the fraction of costs incurred during outage times, solar PV appears in the optimal generation mix. Figure~\ref{fig:cost_reduction_solar_unbounded} displays the installed capacities in an unconstrained scenario for different decarbonization levels and the achieved reduction in total annual system costs of the hybrid BPS including solar PV. It has to be noted that the displayed capacities are not realistically achievable in a majority of real-world applications due to space constraints.

\begin{figure}[!htb]
    \centering
    
    \includegraphics[width= 0.8 \textwidth]{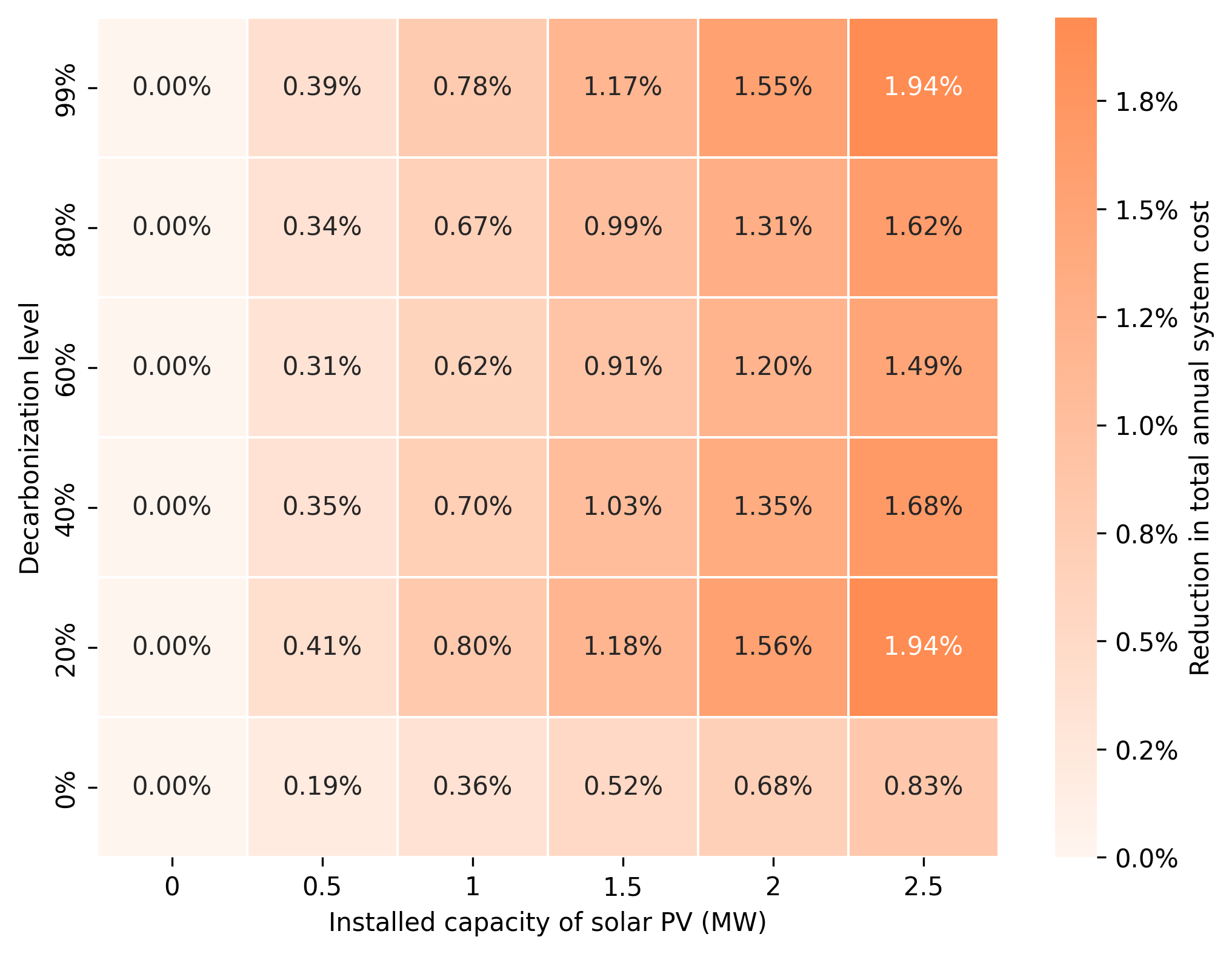}
    \caption{Heatmap of cost reduction potential of solarPV at different decarbonization levels and constrained installed capacities using pro-rata cost accounting.}
    \label{fig:solar_savings}
\end{figure}

To account for the aforementioned spatial constraints, a second assessment with limited solar PV capacity is made. Figure~\ref{fig:solar_savings} shows the cost savings potential at different decarbonization levels and constrained solar PV capacities. It can be seen that the addition of more solar capacity leads to an increased cost savings potential and that cost savings are higher at decarbonization levels above 20\%. However, even with high decarbonization targets, the overall cost reductions remain below 2\% across all scenarios. This is mainly due to the inherent stochastic nature of solar power production. These findings show the limited potential of solar PV for clean BPSs in New England, even under the most favorable cost assumptions.

\subsubsection{Influence of variability in demand}
An evaluation is conducted on the cost-effectiveness and exact composition of the aforementioned hybrid BPSs consisting of clean fuels and batteries across a broader range of applications. The analysis assesses the impact of varying demand patterns on the optimal BPS system. The focus of the analysis is set to variations of the ratio of power demand amplitude to average demand. This ratio is around 1.5 for the examined representative round-the clock medical facility. Ratios between 0, representing constant demand, and 3.5, representing highly fluctuating demand, are analyzed. The total energy consumption remains unchanged across all scenarios, meaning that there are no changes in storage size and that the only variations in total annual system cost are due to different combinations of generating capacities.

\begin{figure}[!htb]
    \centering
    
    \includegraphics[width= 0.8 \textwidth]{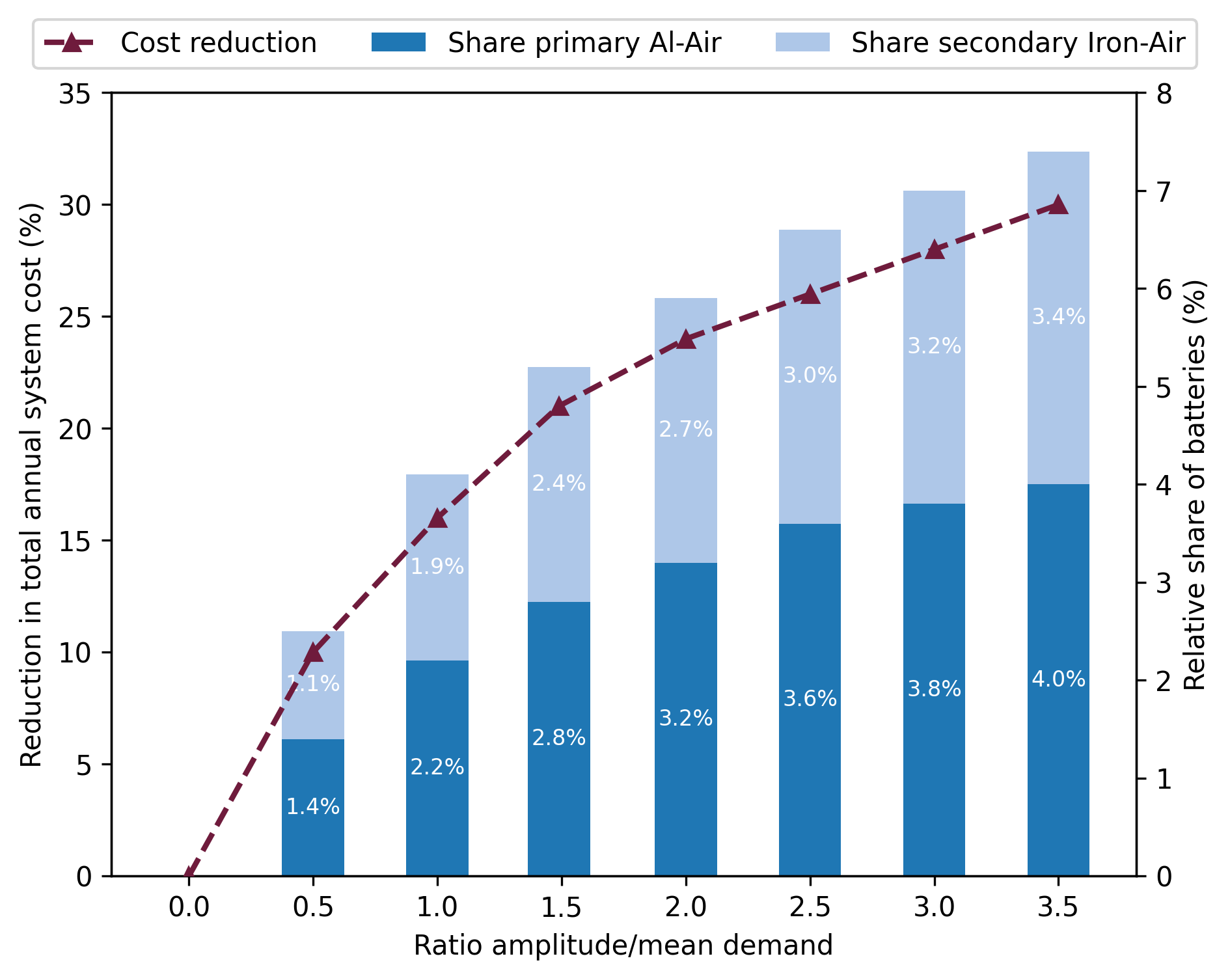}
    \caption{Cost savings and share of primary and secondary batteries relative to demand variability given by the ratio of amplitude to mean demand.}
    \label{fig:demand_variability}
\end{figure}

Figure~\ref{fig:demand_variability} shows the cost savings that can be achieved by an optimally sized hybrid BPS compared to a sizing for peak demand. The results show that the shares of primary and secondary batteries increase for more volatile demand patterns. Simultaneously, cost savings of optimally sized hybrid BPSs also increase for higher demand variability. For the case of a representative round-the-clock medical facility with a 1.5 amplitude/ mean demand ratio, a 21\% cost reduction can be achieved using shares of 2.8\% and 2.4\% of primary Al-air and secondary Iron-air batteries respectively. The observation, that not only the share of secondary but also primary batteries increases for higher demand variability, suggests that primary batteries are indeed cost-effective to cover the most extreme peaks in power demand. This aligns with earlier findings of the study. It is important to highlight that the plot shows the results for a BPS without emission limits, which thus uses a diesel generator as fuel-based technology. According to the aforementioned findings, the results in this plot do thus reflect a lower bound on the share of batteries and cost savings that can be achieved when using other clean fuel-based systems. To conclude, the results show the effectiveness of batteries in covering peak loads and reducing the need for costly scaling of fuel-only-based systems. The integration of BESSs into optimized BPSs can therefore be particularly valuable in applications with high demand variability. 

\subsubsection{Influence of demand response}
Different backup power applications are not only characterized by their variability in power demand but also by their possibility to provide demand side management and demand response. While the default case in this study assumes that demand must be met at each time step during the entire duration of a power outage, some real-world applications offer a certain potential for demand response. Hospitals, for example, can delay non-critical operations and data centers can reschedule certain computing tasks, providing both power and temporal flexibility. The following paragraphs show how the design of optimized BPSs changes with respect to varying demand flexibility. While there exist several methods to modify demand, this study focuses specifically on load shifting and assumes that load shedding is not feasible. For this reason, the entire energy demand during a power outage must be met. It is assumed that shifting demand in time does not result in any additional costs or energy consumption and that demand can be shifted forward and backward in time.

\begin{figure}[H]
\centering
\subfloat[\centering \label{fig:flexible_demand_cost_savings}]{%
    \includegraphics[width=7.7cm]{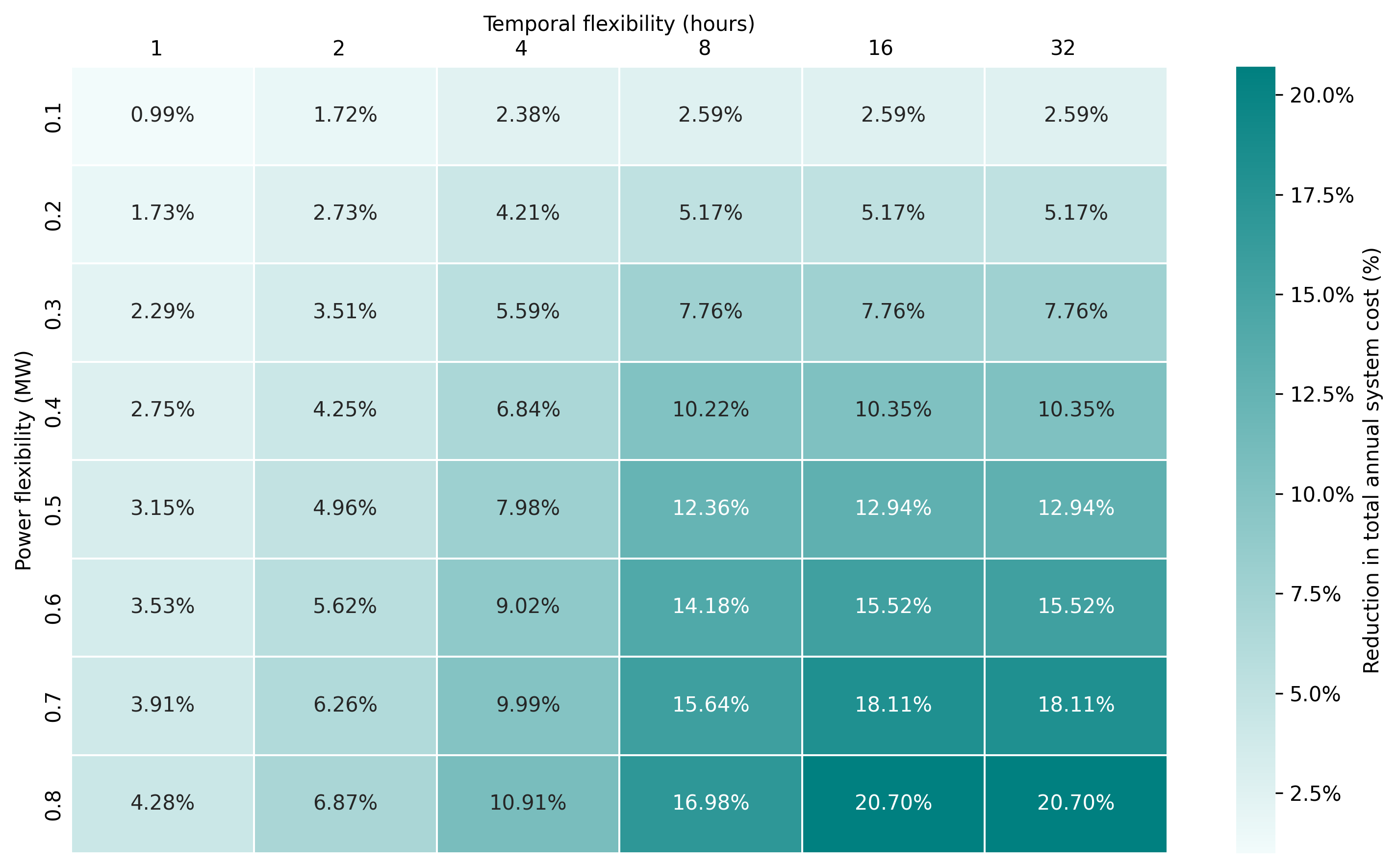}
}
\hfill
\subfloat[\centering \label{fig:flexible_demand_peak_demand}]{%
    \includegraphics[width=7.7cm]{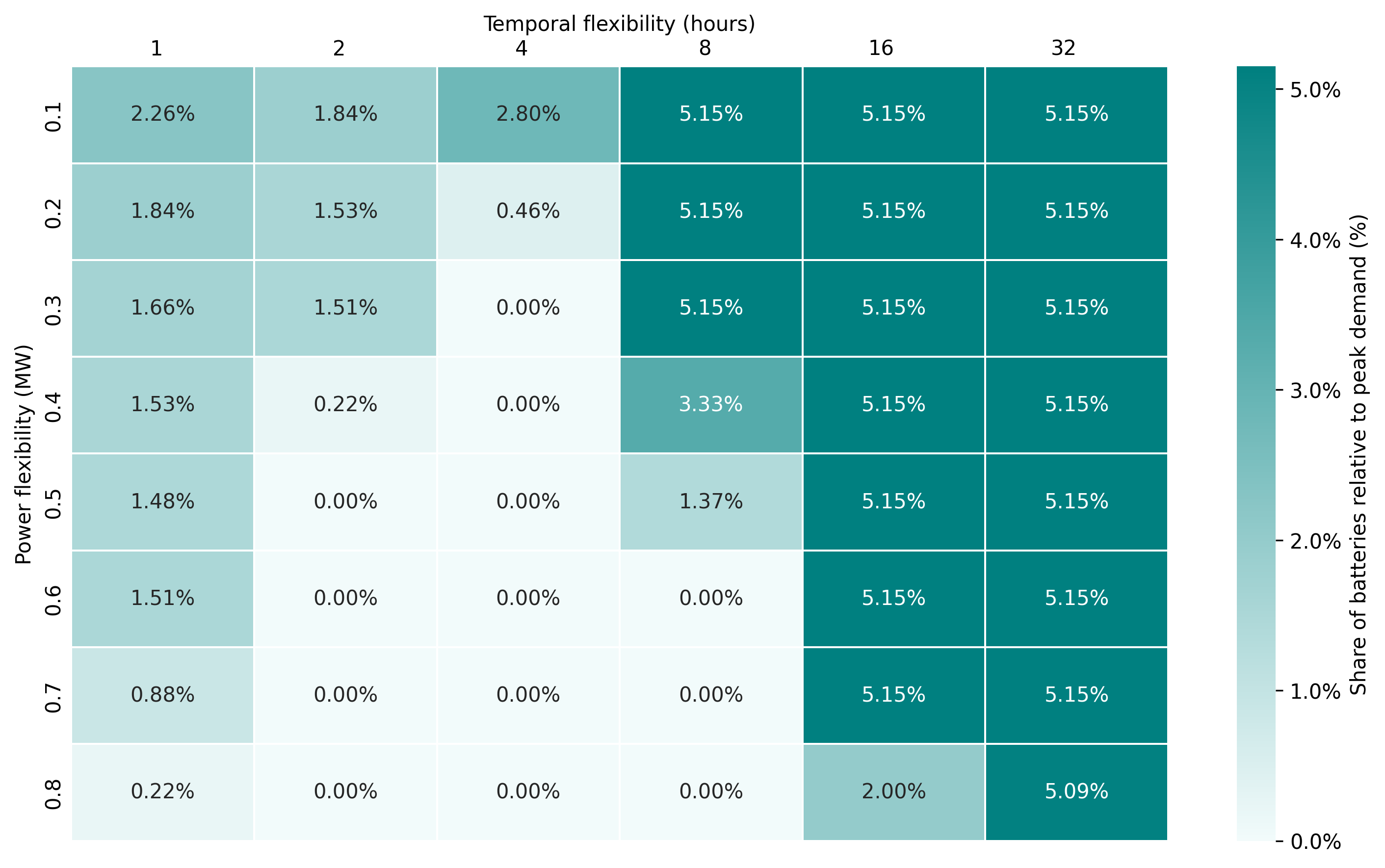}
}
\caption{Impact of demand flexibility on system cost and battery sizing for 2.2 MW demand peak. (\textbf{a}) Reduction in total annual system cost for different levels of temporal and power flexibility. (\textbf{b}) Share of batteries as percentage of peak demand for varying levels of temporal and power flexibility.}
\label{fig:flex_demand_figs}
\end{figure}

Figure~\ref{fig:flexible_demand_cost_savings} shows the impact of demand flexibility on total annual system cost. The temporal flexibility refers to the number of hours ($n$) that a given amount of power at time step $t$ can be shifted forward and backward to all time steps between $t-n$ and $t+n$. The plot shows that both higher power and temporal flexibility lead to decreased total annual system costs of BPSs. Cost reductions can reach up to 20\% for high power and temporal flexibility. The results show that applications with a temporal demand flexibility beyond 16 hours do not benefit from further cost reductions. This is because temporal flexibility beyond a certain value does not enable additional smoothing of demand and that the great majority of power outages is characterized by short durations up to several hours.

The possibility for demand response measures does not only influence the cost savings potential but also leads to changes in the optimal generation mix of BPSs. Figure~\ref{fig:flexible_demand_peak_demand} shows the impact of demand flexibility on the optimal share of BESSs in hybrid BPSs relative to the initial peak demand. The plot shows that the share of batteries decreases for higher power flexibility. The trend is more nuanced for the temporal flexibility component. Increasing the temporal flexibility at low levels first leads to a decrease in the share of batteries. However, at higher temporal flexibilities the share of batteries attains the original value. It is thus cost-efficient to first use demand response to shift peaks in demand and at higher flexibility potentials shift base demand. When shifting base demand the characteristic demand peaks are preserved and cost-effectively covered with batteries. This phenomenon of reintroduction of batteries is observed later for high power flexibilities. To conclude, demand response options partially compete with the integration of batteries in clean fuel-based BPSs. However, batteries can be economically viable even at high demand variability.

\section{Discussion}

While BPSs are often conceptualized as homogeneous systems, our study shows that hybrid BPSs are cost-effective across a wide range of applications and decarbonization scenarios. This involves using primary and secondary batteries as a "cheap power, expensive energy" addition to clean fuel-based systems. The optimal share of batteries in the generation mix positively correlates with power investment costs of generators and fuel cells, which serve as core technologies in hybrid systems. The incentive for BPS operators to invest in hybrid BPSs increases with demand fluctuations and power investment costs of the core technology. Actual cost reduction opportunities may exceed the estimates presented in this study, as oversizing of generators is frequently observed in real-world applications.

Given the infrequent and often unforeseeable nature of BPS usage, it is critical to account for emissions not only from their operation, but also from associated maintenance and testing activities. With our comprehensive techno-economic analysis we address the limited focus on this crucial element in previous research. Accounting for emissions due to fuel replacement and system testing, we identify several decarbonized BPSs. Hybrid BPSs composed of ammonia generators and secondary iron-air batteries offer a cost-effective solution to decrease carbon emissions by up to 79\% compared to EDGs while systems composed of hydrogen fuel cells and secondary iron-air batteries achieve emission reductions of up to 96\%. Standalone BESSs systems are found to be uneconomical for emergency backup systems. Primary al-air batteries can be cost-effective to cover demand peaks and gain more traction in scenarios allowing for emergency fuel purchases, as they allow for cheap power coverage and can be swapped during an outage. However, the overall potential of primary al-air batteries is found to be limited. 

The potential of onsite VRE to complement hybrid BPSs is found to be limited. The findings indicate that investing in solar PV solely for backup power is not economically viable. A best-case scenario using a pro-rate cost attribution for the supply of backup power shows a theoretical potential for cost savings for large solar PV installations. Taking into account space constraints, the realistically achievable cost reduction when integrating solar PV in the optimal generation mix is not substantial. While the cost-efficiency of solar PV for the sole purpose of backup power supply is projected to remain low, the exact potential may vary based on the geographic area. 

Policies have a substantial impact on the costs and emissions of BPSs. The analysis of emergency fuel purchases reveals that both costs and emissions can be significantly reduced when allowing for fuel deliveries during power outages. The main reasons are the decreased upfront investment costs for storage and the reduced maintenance costs associated with fuel replacement and system testing. The results indicate that fuel purchases at 48-hour and 24-hour intervals can reduce total annual system costs by 34\% and 46\%, respectively. Furthermore, these cost savings are found to be largely unaffected by variations in purchase price premiums, making fuel purchases a cost-effective option even with sudden price increases during outages. While most healthcare facilities are legally required to maintain enough fuel onsite to power their operations for 96 hours or more, other applications with BPSs are not suspect to such requirements. For use cases particularly in the high value industry this trade-off between security of supply and adoption of cost-efficient and low-emitting BPSs has to be carefully considered. Operators of multiple sites requiring BPSs could potentially lower costs by maintaining a centralized fuel storage. The introduction of emergency fuel purchases does not change the optimal set of technologies for deployment in decarbonized BPSs systems. Biofuel only becomes a viable option in scenarios with highly volume-constrained fuel deliveries, benefiting from its high volumetric energy density.

It is important to highlight the areas of applicability of the study. While the load profile and power outage data are based on a representative around-the-clock medical facility in Massachusetts, the study's findings apply to a broad range of applications. The evaluation of variations in demand patters, demand response options and sensitivity analysis on key parameters such as the overall power outage likelihood extend the applicability to other critical societal infrastructure and the high-value industry. The findings are therefore applicable to use cases such as laboratories and data centers with a power demand ranging between approximately 500 kW and 50 MW. The results remain relevant at other capacities but scale-dependent cost parameters need to be considered. The findings are applicable in regions with an overall power outage probability between 0.003\% and 1\%.

Some limitations are inherent to the chosen approach and impact the results of the study. The most important constraints and assumptions are outlined in this paragraph and may serve as basis for future research.
\begin{itemize}
  \item Adherence to protocols: The analysis assumes that operators of BPSs follow standard protocols, such as adhering to fuel replacement intervals and frequency of system testing. In practice, these intervals are not always respected, which can lead to changes to the overall cost and emission structure. Accurately capturing user behavior is challenging as it can deviate from the assumed patterns. Combined with irregular system testing intervals, these variations can influence overall system performance.
\item Local VRE availability: The study accounts for variations in demand patterns and different levels of demand flexibility. Regarding the supply side, VRE availability is largely determined by the geographic area under investigation. While solar PV is considered in this study, different VRE resources such as co-located wind turbines may be relevant in other regions.
\item No predictive insight into future outages: The analysis assumes no foresight on timing and duration of power outages. While this assumption is realistic in many scenarios, it can be argued that especially large-scale outages due to extreme weather events are predictable to some extend. Such foresight would enable the operator of a BPS to manage the system more effectively. For example, a secondary battery could be used for other purposes during non-outage times and charged to a high state of charge prior to a predicted outage event. 
\end{itemize}


\section{Conclusions}

This work proposes a scenario-based stochastic optimization framework to evaluate a range of clean technologies for emergency backup power. The analysis incorporates the following steps: (i) it assesses a pool of 27 technologies, including both established and emerging technologies; (ii) it implements a stochastic optimization method, taking into consideration actual demand profiles and outage probabilities; (iii) it evaluates the cost-effectiveness of individual technologies and combined systems; (iv) it highlights crucial factors to achieve deep decarbonization levels, including a thorough emission accounting framework. Acknowledging the diverse applications of BPSs, the study accounts for varying demand patterns and demand response options.

By applying the framework to a real-world setting, we demonstrate the potential of integrating primary and secondary batteries with clean fuel-based systems. Leveraging synergies and complementary cost structures of these technologies significantly reduces total annual system cost. The results underline the cost-efficiency of hybrid BPSs across a wide range of applications and decarbonization levels. Emerging primary batteries are found to be cost-effective for managing demand peaks, though their overall potential is limited. Secondly, the results highlight the importance of accurate emission accounting for BPSs, while also addressing the impact of user compliance and policy frameworks on system performance. The analysis shows that fuel degradation, system testing and maintenance significantly contribute to total system costs and emissions. Protocol adherence as well as policies governing emergency fuel purchases are additional key factors that influence overall system performance. Finally, the exact composition of optimal BPSs varies for different levels of decarbonization. The most cost-effective BPS to achieve substantial emission reduction is composed of ammonia generators paired with secondary iron-air batteries. Hydrogen fuel cells combined with secondary iron-air batteries are the preferred choice for deep decarbonization. Although standalone secondary batteries can theoretically achieve the highest emission reduction under the specified emission accounting framework, they are not economically viable.

In conclusion, the research findings offer broadly applicable and valuable insights to facilitate the widespread adoption of clean backup power solutions. The study can inform decision makers on investment choices for BPSs in sectors such as healthcare, data centers and high value industries. It can also provide policy makers with guidance to effectively drive decarbonization in this sector.



\vspace{6pt}

\subsection*{Acknowledgments} 
The authors extend their sincere gratitude to Daniel Kuhn for his invaluable contributions to this work. Furthermore, the authors acknowledge the MIT Energy Initiative and Princeton University ZERO lab for their contributions to the development of GenX, a configurable power system capacity expansion model for studying low-carbon energy futures. https://github.com/GenXProject/GenX

\begin{spacing}{0.9} 
\bibliographystyle{unsrt}
\bibliography{references.bib}

\begin{thebibliography}{10}

\bibitem{do2023spatiotemporal}
Vivian Do, Heather McBrien, Nina~M Flores, Alexander~J Northrop, Jeffrey Schlegelmilch, Mathew~V Kiang, and Joan~A Casey.
\newblock Spatiotemporal distribution of power outages with climate events and social vulnerability in the usa.
\newblock {\em Nature communications}, 14(1):2470, 2023.

\bibitem{diesel_storage_tanks}
Mark Lehnhof.
\newblock Double wall fuel tanks vs. single wall fuel tanks.
\newblock \url{https://northslopechillers.com/blog/double-wall-fuel-tanks-vs-single-wall-fuel-tanks/}, Accessed: 2024-10-12.

\bibitem{diesel_emissions_true}
Auke Hoekstra.
\newblock Producing gasoline and diesel emits more co2 than we thought.
\newblock \url{https://innovationorigins.com/en/producing-gasoline-and-diesel-emits-more-co2-than-we-thought/}, Accessed: 2024-10-12.

\bibitem{marqusee2021impact}
Jeffrey Marqusee, Sean Ericson, and Don Jenket.
\newblock Impact of emergency diesel generator reliability on microgrids and building-tied systems.
\newblock {\em Applied energy}, 285:116437, 2021.

\bibitem{eoff2007diesel}
David Eoff.
\newblock Diesel generator failures: Lessons taught by hurricanes.
\newblock {\em Power Engineering}, 111(8):68--70, 2007.

\bibitem{yang2024economic}
Shuai Yang, Xiaohan Zhao, Victor Nian, Xueqiang Li, Hailong Li, and Shengchun Liu.
\newblock Economic feasibility of using fuel cells as backup power supply in data centers.
\newblock {\em Sustainable Energy Technologies and Assessments}, 69:103892, 2024.

\bibitem{dimitriou2019fully}
Pavlos Dimitriou and Taku Tsujimura.
\newblock A fully renewable and efficient backup power system with a hydrogen-biodiesel-fueled ic engine.
\newblock {\em Energy Procedia}, 157:1305--1319, 2019.

\bibitem{li2018research}
Jianfeng Li, Dongxiao Niu, Ming Wu, Yongli Wang, Fang Li, and Huanran Dong.
\newblock Research on battery energy storage as backup power in the operation optimization of a regional integrated energy system.
\newblock {\em Energies}, 11(11):2990, 2018.

\bibitem{sanni2021analysis}
Shereefdeen~Oladapo Sanni, Joseph~Yakubu Oricha, Taoheed~Oluwafemi Oyewole, and Femi~Ikotoni Bawonda.
\newblock Analysis of backup power supply for unreliable grid using hybrid solar pv/diesel/biogas system.
\newblock {\em Energy}, 227:120506, 2021.

\bibitem{faraji2019comparative}
Jamal Faraji, Masoud Babaei, Navid Bayati, and Maryam A.~Hejazi.
\newblock A comparative study between traditional backup generator systems and renewable energy based microgrids for power resilience enhancement of a local clinic.
\newblock {\em Electronics}, 8(12):1485, 2019.

\bibitem{elkholy2024techno}
MH~Elkholy, Taghreed Said, Mahmoud Elymany, Tomonobu Senjyu, Mahmoud~M Gamil, Dongran Song, Soichiro Ueda, and Mohammed~Elsayed Lotfy.
\newblock Techno-economic configuration of a hybrid backup system within a microgrid considering vehicle-to-grid technology: A case study of a remote area.
\newblock {\em Energy Conversion and Management}, 301:118032, 2024.

\bibitem{dong2018battery}
Jiaojiao Dong, Lin Zhu, Yu~Su, Yiwei Ma, Yilu Liu, Fred Wang, Leon~M Tolbert, Jim Glass, and Lilian Bruce.
\newblock Battery and backup generator sizing for a resilient microgrid under stochastic extreme events.
\newblock {\em IET Generation, Transmission \& Distribution}, 12(20):4443--4450, 2018.

\bibitem{gatta2018replacing}
Fabio~Massimo Gatta, Alberto Geri, Stefano Lauria, Marco Maccioni, Francesco Palone, Pierluigi Portoghese, Luca Buono, and Andrea Necci.
\newblock Replacing diesel generators with hybrid renewable power plants: Giglio smart island project.
\newblock {\em IEEE Transactions on Industry Applications}, 55(2):1083--1092, 2018.

\bibitem{pathak2024utility}
Geeta Pathak, Bhim Singh, and BK~Panigrahi.
\newblock Utility interactive renewable microgrid with backup power functionality.
\newblock {\em IETE Journal of Research}, 70(4):4333--4344, 2024.

\bibitem{li2020dynamic}
Yaowang Li, Shihong Miao, Xing Luo, Binxin Yin, Ji~Han, and Jihong Wang.
\newblock Dynamic modelling and techno-economic analysis of adiabatic compressed air energy storage for emergency back-up power in supporting microgrid.
\newblock {\em Applied energy}, 261:114448, 2020.

\bibitem{suman2021optimisation}
Gourav~Kumar Suman, Josep~M Guerrero, and Om~Prakash Roy.
\newblock Optimisation of solar/wind/bio-generator/diesel/battery based microgrids for rural areas: A pso-gwo approach.
\newblock {\em Sustainable cities and society}, 67:102723, 2021.

\bibitem{shen2021optimal}
Xin Shen, Zhao Luo, Jun Xiong, Hongzhi Liu, Xin Lv, Taiyang Tan, Jianwei Zhang, Yuting Wang, and Yinghao Dai.
\newblock Optimal hybrid energy storage system planning of community multi-energy system based on two-stage stochastic programming.
\newblock {\em IEEE Access}, 9:61035--61047, 2021.

\bibitem{franco2015robust}
John~F Franco, Marcos~J Rider, and Ruben Romero.
\newblock Robust multi-stage substation expansion planning considering stochastic demand.
\newblock {\em IEEE Transactions on Power Systems}, 31(3):2125--2134, 2015.

\bibitem{jawad2018robust}
Muhammad Jawad, Muhammad~B Qureshi, Muhammad~US Khan, Sahibzada~M Ali, Arshad Mehmood, Bilal Khan, Xiaoyu Wang, and Samee~U Khan.
\newblock A robust optimization technique for energy cost minimization of cloud data centers.
\newblock {\em IEEE Transactions on Cloud Computing}, 9(2):447--460, 2018.

\bibitem{fuel_polishing}
Pristine Diesel.
\newblock Diesel fuel polishing for industrial generators.
\newblock \url{https://pristinediesel.us/}, Accessed: 2024-11-06.

\bibitem{yang2022optimizing}
Haoxiang Yang, Daniel Duque, and David~P Morton.
\newblock Optimizing diesel fuel supply chain operations to mitigate power outages for hurricane relief.
\newblock {\em IISE Transactions}, 54(10):936--949, 2022.

\bibitem{brelsford2024dataset}
Christa Brelsford, Sarah Tennille, Aaron Myers, Supriya Chinthavali, Varisara Tansakul, Matthew Denman, Mark Coletti, Joshua Grant, Sangkeun Lee, Karl Allen, et~al.
\newblock A dataset of recorded electricity outages by united states county 2014--2022.
\newblock {\em Scientific Data}, 11(1):271, 2024.

\bibitem{ikotun2023k}
Abiodun~M Ikotun, Absalom~E Ezugwu, Laith Abualigah, Belal Abuhaija, and Jia Heming.
\newblock K-means clustering algorithms: A comprehensive review, variants analysis, and advances in the era of big data.
\newblock {\em Information Sciences}, 622:178--210, 2023.

\bibitem{na2010research}
Shi Na, Liu Xumin, and Guan Yong.
\newblock Research on k-means clustering algorithm: An improved k-means clustering algorithm.
\newblock In {\em 2010 Third International Symposium on intelligent information technology and security informatics}, pages 63--67. Ieee, 2010.

\bibitem{yuan2019research}
Chunhui Yuan and Haitao Yang.
\newblock Research on k-value selection method of k-means clustering algorithm.
\newblock {\em J}, 2(2):226--235, 2019.

\bibitem{chong2021k}
Bao Chong et~al.
\newblock K-means clustering algorithm: a brief review.
\newblock {\em vol}, 4:37--40, 2021.

\bibitem{gan2017k}
Guojun Gan and Michael Kwok-Po Ng.
\newblock K-means clustering with outlier removal.
\newblock {\em Pattern Recognition Letters}, 90:8--14, 2017.

\bibitem{ruiz2015robust}
Carlos Ruiz and Antonio~J Conejo.
\newblock Robust transmission expansion planning.
\newblock {\em European Journal of Operational Research}, 242(2):390--401, 2015.

\bibitem{outages_massachusetts}
Claes Mats.
\newblock Data study reveals 10 states with the most \& least reliable power grids.
\newblock \url{https://generatordecision.com/states-with-the-most-least-reliable-power-grids/#10-states-with-the-least-reliable-power-grids}, Accessed: 2024-11-01.

\bibitem{genx_source}
MIT~Energy Initiative and Princeton University~ZERO lab.
\newblock A configurable power system capacity expansion model for studying low-carbon energy futures.
\newblock \url{https://github.com/GenXProject/GenX}, Accessed: 2024-11-01.

\bibitem{causes_power_outages}
Constellation.
\newblock 10 common causes of power outages.
\newblock \url{https://www.constellation.com/energy-101/weather-preparedness/causes-of-power-outages.html}, Accessed: 2024-11-03.

\bibitem{outage_likelihood_overall_1}
Casinobonusca.
\newblock Odds of experiencing a power outage today.
\newblock \url{https://casinobonusca.com/odds-of-experiencing-a-power-outage-today/}, Accessed: 2024-11-15.

\bibitem{outage_likelihood_overall_2}
Mats Claes.
\newblock Data study reveals 10 states with the most \& least reliable power grids.
\newblock \url{https://generatordecision.com/states-with-the-most-least-reliable-power-grids/#10-states-with-the-least-reliable-power-grids}, Accessed: 2024-11-15.

\bibitem{genset_sizing}
General Power.
\newblock How to calculate what size generator you need.
\newblock \url{https://www.genpowerusa.com/blog/how-to-calculate-commercial-generator-size}, Accessed: 2024-11-04.

\bibitem{genset_sizing_2}
Generac.
\newblock Generator sizing guide.
\newblock \url{https://www.homepowersystems.net/wp-content/uploads/imported/pdf/GeneratorSizingGuide.pdf}, Accessed: 2024-11-04.

\bibitem{tol2023social}
Richard~SJ Tol.
\newblock Social cost of carbon estimates have increased over time.
\newblock {\em Nature climate change}, 13(6):532--536, 2023.

\end{thebibliography}


\begin{thebibliography}{999}
\bibitem[Aranceta-Bartrina(1999a)]{ref-journal}
Aranceta-Bartrina, Javier. 1999a. Title of the cited article. \textit{Journal Title} 6: 100--10.
\bibitem[Aranceta-Bartrina(1999b)]{ref-book1}
Aranceta-Bartrina, Javier. 1999b. Title of the chapter. In \textit{Book Title}, 2nd ed. Edited by Editor 1 and Editor 2. Publication place: Publisher, vol. 3, pp. 54–96.
\bibitem[Baranwal and Munteanu {[1921]}(1955)]{ref-book2}
Baranwal, Ajay K., and Costea Munteanu. 1955. \textit{Book Title}. Publication place: Publisher, pp. 154--96. First published 1921 (op-tional).
\bibitem[Berry and Smith(1999)]{ref-thesis}
Berry, Evan, and Amy M. Smith. 1999. Title of Thesis. Level of Thesis, Degree-Granting University, City, Country. Identifi-cation information (if available).
\bibitem[Cojocaru et al.(1999)]{ref-unpublish}
Cojocaru, Ludmila, Dragos Constatin Sanda, and Eun Kyeong Yun. 1999. Title of Unpublished Work. \textit{Journal Title}, phrase indicating stage of publication.
\bibitem[Driver et al.(2000)]{ref-proceeding}
Driver, John P., Steffen Rohrs, and Sean Meighoo. 2000. Title of Presentation. In \textit{Title of the Collected Work} (if available). Paper presented at Name of the Conference, Location of Conference, Date of Conference.
\bibitem[Harwood(2008)]{ref-url}
Harwood, John. 2008. Title of the cited article. Available online: URL (accessed on Day Month Year).
\end{thebibliography}
\end{spacing}

%


\end{document}